\documentclass[12pt]{aastex}

\usepackage{emulateapj5}
\usepackage{psfig}

\newcommand{\flam}{ergs~cm$^{-2}$~s$^{-1}$~\AA$^{-1}$}
\newcommand{\msun}{\mbox{$\:M_{\sun}$}}

\newcommand{\expu}[3]{\mbox{\rm $#1 \times 10^{#2} \rm\:#3$}}

\begin{document}

\submitted{To appear in the Astrophysical Journal.}

\title{FUSE Observations of U Geminorum During Outburst and
Decline
}

\author{Cynthia S.\ Froning, Knox S.\ Long}
\email{froning@stsci.edu, long@stsci.edu}
\affil{Space Telescope Science Institute, \\ 3700 San Martin Drive,
Baltimore, MD 21218}
\author{Janet E.\ Drew}
\email{j.drew@ic.ac.uk}
\affil{Imperial College of Science, Technology and Medicine, \\
Blackett Laboratory, Prince Consort Rd., \\ London SW7 2BZ UK}
\author{Christian Knigge}
\email{christian@astro.soton.ac.uk}
\affil{Department of Physics \& Astronomy,\\ University of Southampton, \\
Southampton SO17 1BJ UK}
\and
\author{Daniel Proga}
\email{proga@sobolev.gsfc.nasa.gov}
\affil{Laboratory for High Energy Astrophysics, \\ NASA Goddard Space Flight
Center, \\ Greenbelt, MD 20771 USA}

\begin{abstract}

We have obtained FUV (904 -- 1187~\AA) spectra of U~Gem in outburst
with FUSE.  Three of the observations were acquired during the plateau
phase of the outburst, while the fourth was obtained during late
outburst decline.  The plateau spectra have continuum shapes and
fluxes that are approximated by steady-state accretion disk model
spectra with $\dot{m} \simeq$ \expu{7}{-9}{\msun \: yr^{-1}}.  The
spectra also show numerous absorption lines of \ion{H}{1},
\ion{He}{2}, and two- to five-times ionized transitions of C, N, O, P,
S and Si. There are no emission features in the spectra, with the
possible exception of a weak feature on the red wing of the \ion{O}{6}
doublet.  The absorption lines are narrow (FWHM $\sim$
500~km~s$^{-1}$), too narrow to arise from the disk photosphere, and
at low velocities ($\leq$700~km~s$^{-1}$). The \ion{S}{6} and
\ion{O}{6} doublets are optically thick.  The absorption lines in the
plateau spectra show orbital variability: in spectra obtained at
orbital phases $0.53 \leq \Phi \leq 0.79$, low-ionization absorption
lines appear and the central depths of the pre-existing lines
increase.  The increase in line absorption occurs at the same orbital
phases as previously observed EUV and X-ray light curve dips.  If the
absorbing material is in (near-) Keplerian rotation around the disk,
it must be located at large disk radii. The final observation occurred
when U~Gem was about two magnitudes from optical quiescence.  The
spectra are dominated by emission from an $\simeq$43,000~K,
metal-enriched white dwarf (WD).  The inferred radius of the WD is
\expu{4.95}{8}{cm}, close to that observed in quiescence. Allowing for
a hot heated region on the surface of the WD improves the fit to the
spectrum at short ($<$960~\AA) wavelengths.
\end{abstract}


\section{Introduction} \label{sec_intro}

Cataclysmic variables (CVs) are mass-exchanging binary systems in
which a late-type donor star transfers mass to a white dwarf (WD).
Dwarf novae (DN) are CVs that undergo semi-regular outbursts,
brightening by 3 -- 5 visual magnitudes.  The outbursts last from
about a day to several weeks, depending on the system.  Mass exchange
in DN is regulated by an accretion disk around the WD.  The outbursts
are triggered by a thermal instability in the accretion disk that
drives the disk from a low temperature, low mass accretion rate
($\dot{m}$) quiescent state to a hot, high-$\dot{m}$ outburst state.
The properties of CVs, including DN, have been reviewed in detail by
\citet{warner1995}.

In the ``standard'' theory of disk accretion in DN (see, e.g.,
Lynden-Bell \& Pringle 1974), half of the accretion energy is radiated
away by the disk and half by a boundary layer (BL) between the disk
and the slowly- or non-rotating WD.  In quiescence, when the mass
accretion rate is low, the disk is believed to be cool (T$_{disk} <$
8000~K) while the boundary layer is very hot (T$_{BL} \sim 10^{8}$~K)
and optically thin.  Neither is expected to contribute substantially
to the ultraviolet (UV) flux, which instead should be dominated by
emission from the WD.  In outburst, the disk and the BL become
optically thick, and the accretion disk is expected to be the
principal UV emitter (T$_{disk} \sim 30,000$~K and T$_{BL} \sim
10^{5}$~K).  In addition to continuum emission from the thermalized
disk, many CVs, including novalikes and DN in outburst, show
blueshifted UV absorption that indicates the presence of strong, fast
winds from the disk or the WD (e.g., Heap et al.\ 1978; Holm, Panek \&
Schiffer 1982; C\'{o}rdova \& Mason 1982).

The first CV to be discovered, U~Gem remains the prototypical DN and
among the best studied.  It goes into outburst about every 118 days
and brightens by 5 magnitudes; the outbursts have a mean duration of
12 days \citep{szkody1984}. U~Gem has an orbital period of 4.25~hrs;
the masses of the WD ($\sim$1.1~\msun), the mass donor star
($\sim$0.4~\msun), and the inclination ($\sim67\arcdeg$) are well
determined \citep{smak1976,sion1998,long1999}. Its distance
(96.4$\pm$4.6~pc) is known accurately as a result of astrometric
determination with the FGS on HST \citep{harrison1999}.  U~Gem system
parameters adopted for this study are given in Table~\ref{tab_par}.

In most respects, U~Gem's observational characteristics hew to the
simple theoretical picture for DN outlined above.  In outburst, the UV
spectrum is consistent with that expected from a steady-state
accretion disk with $\dot{m}$ $\sim$ \expu{5}{17}{g \: s^{-1}}
(\expu{8}{-9}{\msun \: yr^{-1}}; Panek \& Holm 1984).  In quiescence,
its UV spectrum is well fit by WD models with a temperature of
$\sim$30,000~K far from outburst \citep{panek1984,long1993}.  Its
X-ray/EUV outburst spectrum is soft, with a characteristic temperature
of 140,000~K \citep{cordova1984a,cordova1984b,long1996}, while in
quiescence the X-ray spectrum hardens and can be fit in terms of a
thermal plasma with a temperature of 5 -- 10~keV (\expu{6}{7}{K} --
\expu{1}{8}{K}; Szkody et al.\ 1996).  Furthermore, in U~Gem there is
observational verification that the BL and accretion disk luminosities
are comparable in outburst \citep{long1996}.

U~Gem's X-ray and EUV light curves show dips around orbital phase 0.7
akin to the X-ray ``dippers'' in low mass X-ray binaries.  These dips
indicate the presence of absorbing material located well above the
disk plane and are seen in both outburst and quiescence
\citep{mason1988,long1996,szkody1996}.  Other interesting
characteristics of U~Gem include the fact that the WD appears to cool
from an average temperature of 38,000~K shortly after outburst to
30,000~K several months later \citep{kiplinger1991,long1994a}. The UV
flux does not decline as much as expected for this drop in
temperature.  Long et al.\ (1993) suggested that an ``accretion belt''
may exist on the WD, a region hotter than the bulk WD temperature; it
is this belt, rather than the entire WD, that cools during quiescence
(see also Cheng et al.\ 1997). Models of the WD spectrum in U~Gem have
also indicated WD surface abundances that show evidence of CNO
processing \citep{sion1998,long1999}.

Because U~Gem's observational behavior tallies well with the basic
theoretical underpinnings of DN accretion, U~Gem is an excellent
target for in-depth analyses of the physics of accretion in disk-fed
systems and the evolution of DN phenomenology during the outburst
cycle.  The far-ultraviolet (FUV) provides an excellent viewing window
for such analyses.  The continuum FUV in outburst is dominated by
emission from the innermost regions of the accretion disk, near the
WD, while the numerous FUV spectral lines are probes of the
temperature and ionization structure of the disk, the winds, and other
regions in the binary, such as the source of the X-ray absorption
dips.

We have obtained FUV spectra of U~Gem at the peak and through the
decline of its 2000 March outburst using the Far Ultraviolet
Spectroscopic Explorer (FUSE; Moos et al.\ 2000).  The spectra cover a
904 -- 1187~\AA\ wavelength range at spectral resolutions of
$\sim$12,000.  We report here the results of those observations.  The
observation log and our approach to reducing the data are described in
\S\ref{sec_redn}.  The basic characteristics of the FUV spectra are
described in \S\ref{sec_morph}.  Analysis of the continuum and lines
in the spectra from outburst plateau are addressed in
\S\ref{sec_obs123}, while the spectrum in late outburst decline is
analyzed in \S\ref{sec_obs4}.  Discussion of the results and
concluding remarks are presented in \S\ref{sec_disc} and
\S\ref{sec_conc}.

\section{The Observations and their Reduction} \label{sec_redn}

\begin{figure*}
\plotone{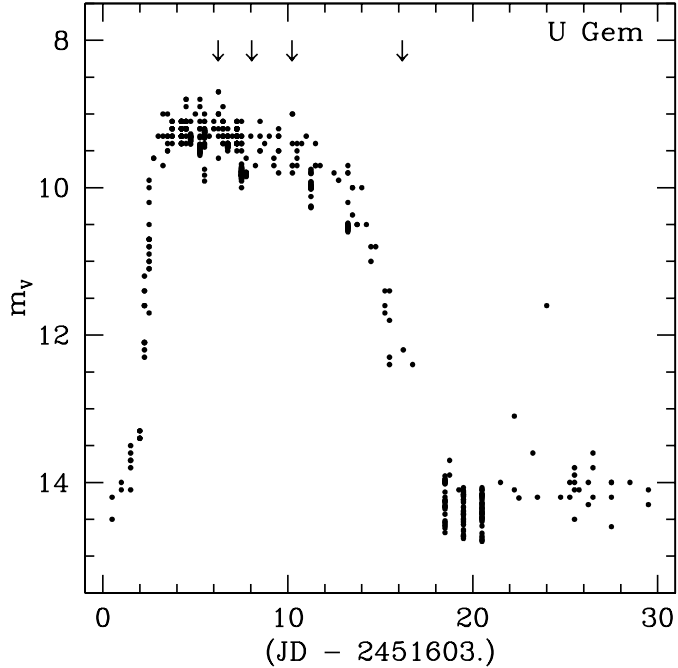}
\figcaption[f1.eps]{The optical light curve of U~Gem during
its 2000 March outburst as reported by the AAVSO.  The times of the
FUSE observations are indicated by the downward arrows. \label{fig_aavso}}
\end{figure*}

We observed U~Gem in the LWRS ($30\arcsec\times30\arcsec$) aperture of
FUSE on 2000 March 5, 7, 9 and 17.  An observation log, including
the binary orbital phases observed, is given in Table~\ref{tab_obs}.
Thanks to prompt notification by the AAVSO and deft scheduling by the
FUSE operations team, we were able to acquire the first spectrum of
U~Gem about 1.5 days after U~Gem reached outburst maximum.  The
relationship between the FUSE observations and the optical light curve
reported by the AAVSO is shown in Figure~\ref{fig_aavso}.  The first
observation was obtained at peak visual brightness of the outburst;
the second and third were obtained during the outburst plateau; and
the last observation was obtained near the end of the outburst, about
2 days before U~Gem returned to its quiescent visual magnitude.  The
full outburst lasted 18 days.

The FUSE telescope collects data simultaneously in four optical
channels.  Each channel covers a portion of the FUSE wavelength range
(905 -- 1187 \AA).  The data are written to eight segment spectra that
partially overlap in coverage and, taken together, cover the full
wavelength range with some redundancy.  Our first three observations
were acquired in the ``spectral image'' mode, since U~Gem was
projected to be too bright for ``time-tag'' observations.  The final
observation, which took place when U~Gem was known to be fading, was
acquired in time-tag mode.  The FUSE instrument and its on-orbit
performance are described in detail by \citet{sahnow2000} and in the
FUSE Observers Guide.

The Obs.\ 1 -- Obs.\ 3 data products consisted of multiple flux
calibrated spectra with 500 -- 700 sec exposure times, spaced
irregularly over the binary orbit.  We used spectra processed by the
FUSE data reduction pipeline (CALFUSE V.\ 1.6.8), but updated the
calibration solutions with efficiency curves and wavelength solutions
from the pipeline V.\ 1.8.7.  The number of spectra (not including their
division into segments) and their exposure times in each of Obs.\ 1 --
Obs.\ 3 are noted in the comments to Table~\ref{tab_obs}.

For each of the Obs.\ 1 -- Obs.\ 3 exposures, we created continuous
spectra by ``stitching'' the data from the eight segments into one.
We made each combined spectrum by specifying a linear (0.07~\AA)
dispersion for the output spectrum and averaging all input data that
fell within a given output wavelength bin.  Each input datum was
weighted by the sensitivity of its pixel of origin.  We masked out
data for which the flux calibration was uncertain --- specifically,
data affected by the ``worm'', an instrumental artifact that disrupts
the LiF1B segment for $\lambda > 1150$~\AA; see the FUSE Data Handbook
v.\ 1.1 for more information.  Depending on the placement of the
target in each of the four channel apertures, there can be offsets in
the absolute wavelength zero points among data from different
channels.  We corrected for the offset in our observations before
averaging by shifting data from the two SiC channels by 0.15~\AA\ to
align them with the LiF data.

\begin{figure*}
\psfig{file=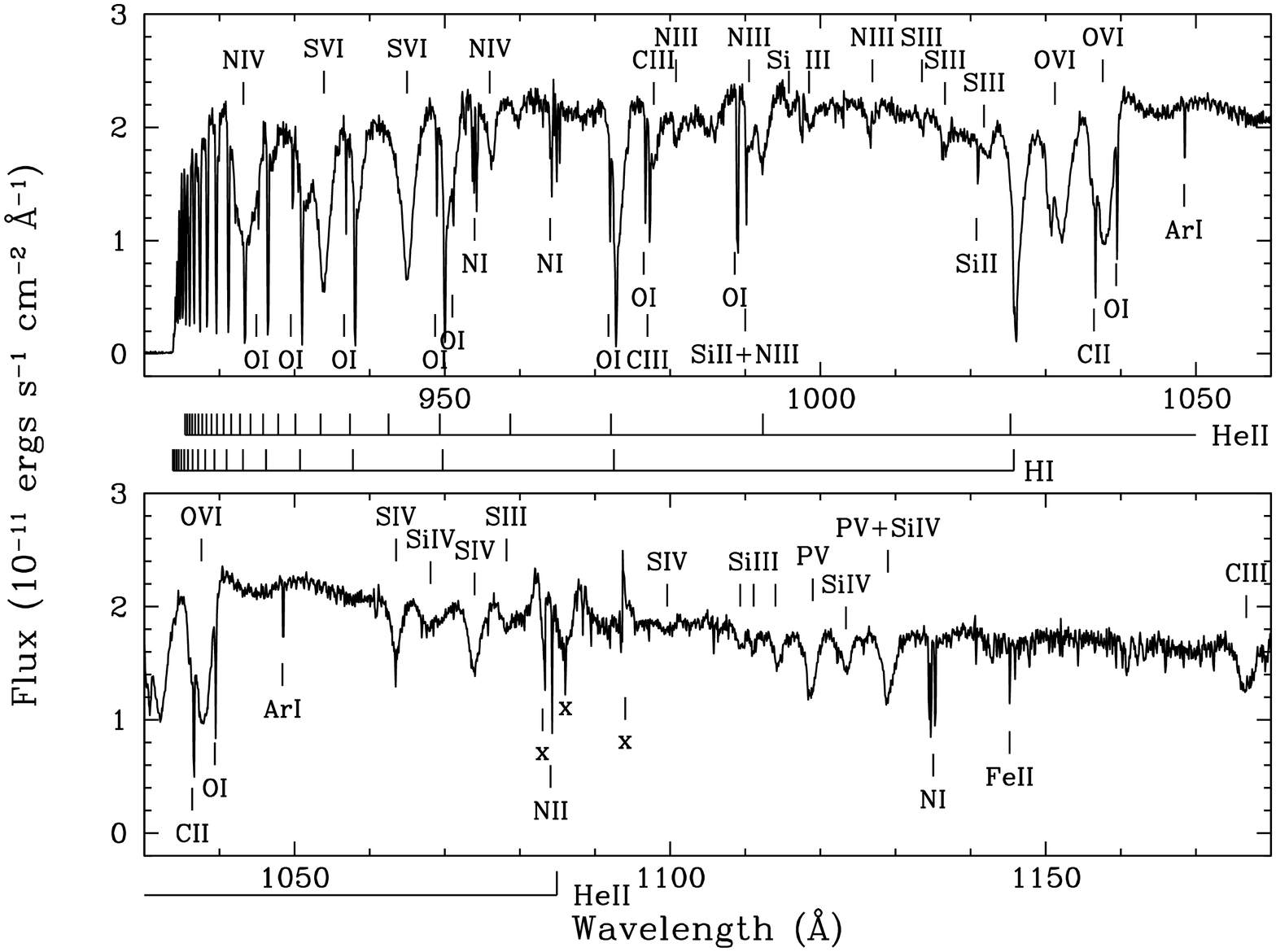,height=5in,angle=-90} \figcaption[f2.eps]{The Obs.\
1 time-averaged FUV spectrum of U~Gem, acquired at the peak of the
2000 March outburst.  The locations of \protect\ion{H}{1} and
\protect\ion{He}{2} lines are labelled below each frame. Absorption
lines of metals intrinsic to U~Gem are labelled above the spectrum,
while prominent interstellar metal lines are labelled below the
spectrum.  The sharp discontinuities at 1070 \protect\AA\ and 1090
\protect\AA\ are artifacts of combining the separate segment spectra
and are not real spectral features; these and other spurious features
are marked with an ``x''.  The \protect\ion{H}{1} transitions are
labelled to the shortest (IS) absorption feature that can be
individually resolved in the spectrum, the 1-21 transition of
\protect\ion{H}{1}, $\lambda$913.826.  The \protect\ion{He}{2}
transitions are arbitrarily labelled to the 2-30 transition,
\protect\ion{He}{2} $\lambda$915.425.  \label{fig_d1}}
\end{figure*}

\begin{figure*}
\psfig{file=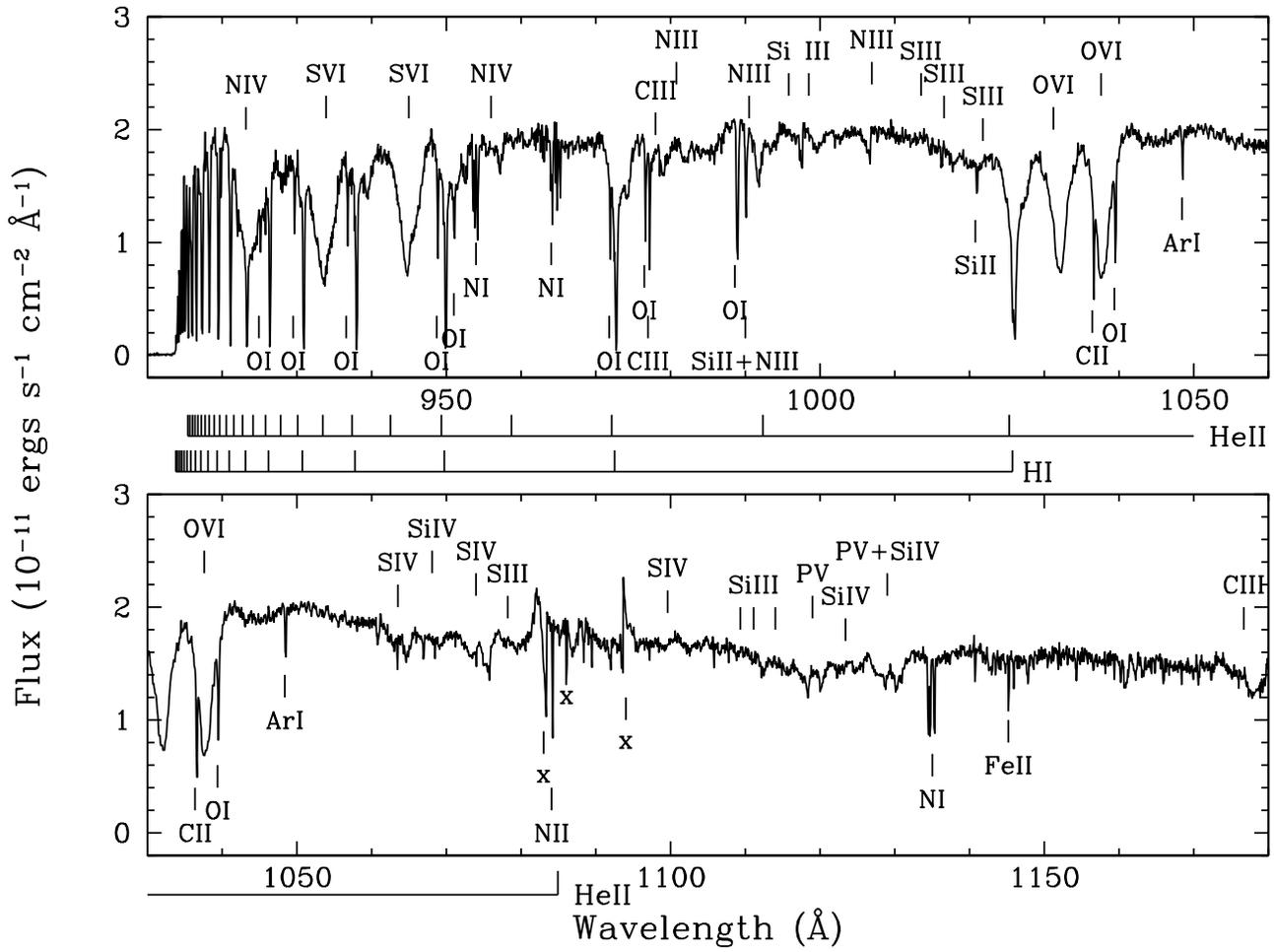,height=5in,angle=-90}
\figcaption[f3.eps]{The Obs.\ 2 time-averaged FUV spectrum of
U~Gem, acquired on the plateau of the 2000 March outburst.
\label{fig_d2}}
\end{figure*}

\begin{figure*}
\psfig{file=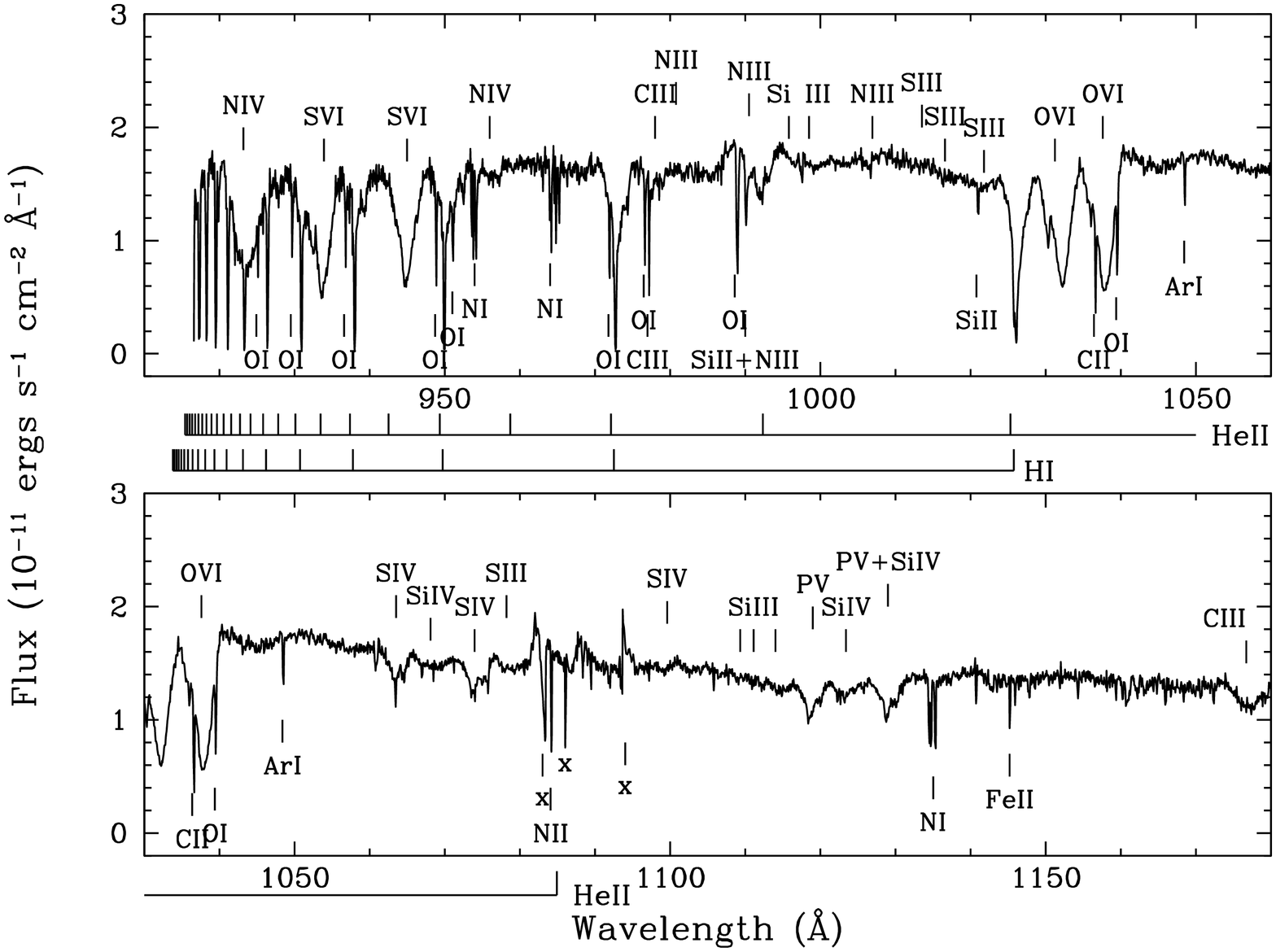,height=5in,angle=-90}
\figcaption[f4.eps]{The Obs.\ 3 time-averaged spectrum of
U~Gem, acquired on the outburst plateau.  The lack of data shortward
of 915~\AA\ is the result of drift in one of the FUSE optical channels
during the observation.  \label{fig_d3}}
\end{figure*}

We also excluded data from the mean spectra that were affected by
channel drift. FUSE guides on only one of its four channels (LiF1,
which covers 987 -- 1082~\AA\ and 1094 -- 1188~\AA).
Thermally-induced motions of the channel mirrors can lead to problems
with co-alignment of all four channels during some observations,
causing loss of data.  Channel misalignments worsened over the course
of our program: while none of the channels were misaligned during
Obs.\ 1, all three non-guided channels drifted during Obs.\ 4.  Any
segment that showed evidence of channel drift --- based on the target
flux in that channel relative to the flux in channel LiF1 over the
region of overlapping wavelength coverage --- was discarded, leading
to some loss of data and wavelength coverage, particularly below
1000~\AA, in the Obs.\ 2 -- Obs.\ 4 spectra.  Finally, we combined all
the exposures from each observation to create time-averaged spectra.
The time-averaged spectra of Obs.\ 1 -- Obs.\ 3 are shown in
Figures~\ref{fig_d1} --~\ref{fig_d3}.  Based on the FWHM of the
interstellar lines, we estimate that the spectral resolution of our
combined segment spectra is $\sim$0.1~\AA.

We reduced the Obs.\ 4 data using a slightly different procedure.
Since the data were acquired in time-tag mode, we were able to correct
for channel drift, which was severe, by directly eliminating data
acquired when a channel was not properly aligned.  We started with the
raw, photon-counting data and re-extracted the spectra (using CALFUSE
V.\ 1.8.7) in 300 sec exposures. We determined which of the segment
spectra were usable by examining the raw count rate light curves in
each channel, and we combined the good segments to create continuous
spectra at 0.07~\AA\ dispersion.  As with the Obs.\ 1 -- Obs.\ 3 data,
we weighted the means by effective area, masked out the ``worm'', and
offset the SiC channel wavelengths by 0.15~\AA\ before combining.

\begin{figure*}
\psfig{file=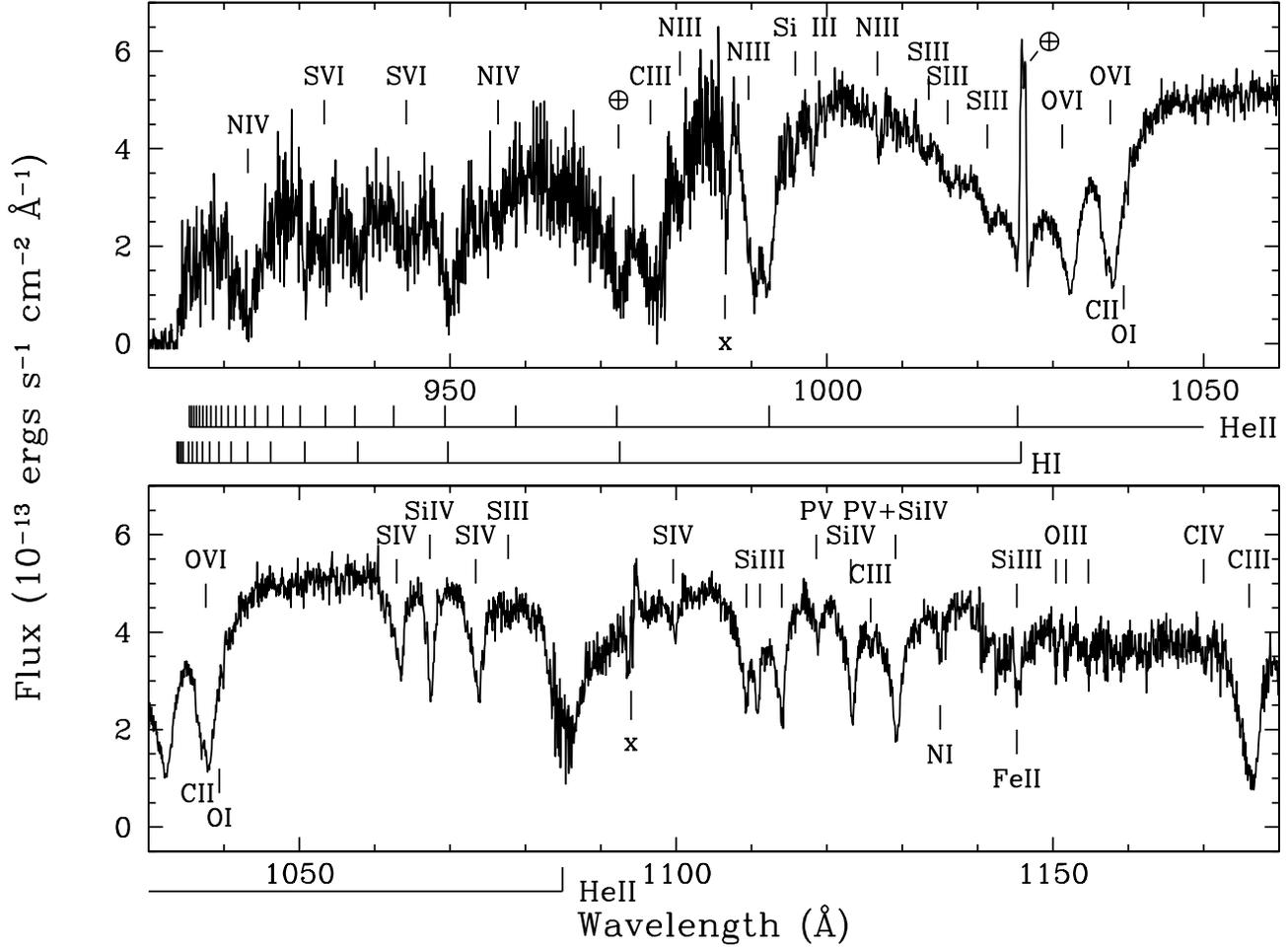,height=5in,angle=-90}
\figcaption[f5.eps]{The Obs.\ 4 time-averaged spectrum of U~Gem,
acquired when the system was at m$_{V} \sim 12$, about 2 days before
the return to optical quiescence. The time-averaged spectrum is the
mean of 42 (300 sec exposure time) spectra.  The 42 spectra were
shifted to remove the orbital motion of the WD before being combined.
The circled crosses indicate lines of terrestrial airglow.  The
airglow line at 972~\AA\ and the interstellar lines have been smeared
out by the WD orbital motion correction.  The x's mark the locations
of spurious features, artifacts of the combination of the original
segment spectra. \label{fig_d4}}
\end{figure*}

The Obs.\ 4 spectra are much fainter than those of Obs.\ 1 -- Obs.\ 3,
and are clearly dominated by the WD (see \S\ref{sec_morph} and
\S\ref{sec_obs4}). Therefore, before creating a time-averaged
spectrum of Obs.\ 4, we corrected for the smearing of the WD spectrum
caused by the orbital radial velocity variations of the WD.  The
orbital period and the radial velocity curve of the WD in U~Gem are
known to high accuracy \citep{marsh1990,long1999}.  Using these, we
shifted each of the 300 sec spectra in wavelength to place them at a
common velocity, the $\gamma$-velocity of the WD.  We then averaged
the shifted spectra to create the time-averaged spectrum of Obs.\ 4
shown in Figure~\ref{fig_d4}.  We did not mask out non-moving features
from interstellar absorption and airglow when creating the
time-averaged spectrum, so these lines have been smeared out in the
spectrum shown in Figure~\ref{fig_d4}.

\section{The Spectra} \label{sec_morph}

The time-averaged spectra of Obs.\ 1 -- Obs.\ 3, acquired during peak
and plateau of the optical outburst, look similar.  Specifically,
the continua rise from zero at the Lyman limit to a peak flux near
1005~\AA\ and decline slowly at longer wavelengths.  The spectra show
a wealth of absorption features.  With the exception of a possible
emission wing on the long wavelength side of the \ion{O}{6} doublet,
there are no emission lines in the spectra. (\ion{He}{2}
$\lambda$1084~\AA\ shows an apparent emission feature, but it is a
data reduction artifact: poor flux calibration at the edges of three
segments led to spurious increases in the flux around this line.)

The spectra also show numerous very narrow (FWHM $\sim$25 km~s$^{-1}$)
absorption lines.  Their widths are of order the estimated resolution
of FUSE LWRS spectra (20 -- 25 km~s$^{-1}$), which suggests that these
lines are unresolved (with the exception of the narrow \ion{H}{1}
lines, which have FWHM $\sim$40 km~s$^{-1}$).  These narrow absorption
lines do not vary in velocity or equivalent width over the
observations.  The same transitions have been observed on lines of
sight to other stars (e.g., Morton 1978), all of which indicates that
the lines are purely interstellar.  Unlike the majority of FUSE
spectra \citep{shull2000} and consistent with its proximity,
interstellar molecular hydrogen absorption features do not appear in
the spectra of U~Gem.

The absorption lines intrinsic to U~Gem (i.e., not interstellar) in
Obs.\ 1 -- Obs.\ 3 include transitions of \ion{H}{1}, \ion{He}{2}, and
numerous metal lines in a range of ionization states, including lines
of \ion{C}{3}, \ion{N}{3}, \ion{Si}{3}, \ion{S}{3}, \ion{N}{4},
\ion{Si}{4}, \ion{S}{4}, \ion{O}{6} and \ion{S}{6}.  The \ion{H}{1}
lines comprise the Lyman series from Ly~$\beta\ \lambda 1025$~\AA\ in
the red to at least the 1-9 transition of \ion{H}{1} at
$\lambda$923.15~\AA, which is the last \ion{H}{1} line that can be
individually distinguished.  There is evidence that \ion{H}{1} line
absorption continues all the way to the Lyman limit, however, in that
the spectra vary at $<$920~\AA\ in the same way that the spectral
lines, but not the continuum, vary at the longer wavelengths (see
\S\ref{sec_lines}).  The \ion{He}{2} lines in the spectra include
\ion{He}{2} $\lambda$1084.9~\AA, $\lambda$992.4~\AA\ and
$\lambda$958.7~\AA.  Two other \ion{He}{2} absorption lines,
$\lambda$972.1~\AA\ and $\lambda$1025.3~\AA, are probably present but
are blended with \ion{H}{1}.  Unlike the \ion{H}{1} absorption, the
\ion{He}{2} absorption does not persist to the shortest wavelengths,
as the \ion{He}{2} lines at $\lambda$942.5~\AA\ and
$\lambda$927.9~\AA, unblended with \ion{H}{1}, are not seen in the
spectra.

Based on low-order spline fits to the line-free regions of the
time-averaged spectra, the FUV continuum dropped by 8\% -- 10\%
between Obs.\ 1 and Obs.\ 2 and 12\% -- 14\% between Obs.\ 2 and Obs.\
3.  The peak flux (at $\lambda$1005~\AA) is \expu{1.9}{-11}{$\flam$}
in Obs.\ 1, \expu{1.75}{-11}{$\flam$} in Obs.\ 2 and
\expu{1.5}{-11}{$\flam$} in Obs.\ 3.  (The uncertainty in the FUSE
flux calibration is estimated at 15\%; FUSE Data Handbook, V.\ 1.1.)
The shape of the continuum is the same in all three observations: the
continuum decline was gray.  The continuum does not vary much within a
given observation.  To quantify this, we examined spectra from the
LiF1 channel.  This is the guide channel for FUSE, and its spectra are
unaffected by changes in channel alignment.  Measurements of the mean
continuum flux in a line-free region (1045 -- 1055~\AA) of the spectra
indicate little continuum variation: none of the 500 -- 700 sec
spectra in each observation are more than 3\% from the overall
observation mean, and the standard deviation of the individual spectra
means about the overall mean is $\leq$2\%.

U~Gem had faded at UV wavelengths by Obs.\ 4. The time-averaged
spectrum peaks at 1000~\AA\ at \expu{5}{-13}{$\flam$}, a factor of 30
below the Obs.\ 3 peak flux.  Low-ionization lines of
\ion{He}{2}, \ion{C}{3}, \ion{N}{3} and \ion{Si}{3}, as well as
\ion{Si}{4}, are more prominent in Obs.\ 4 than in the previous
observations.  In addition, a number of low-ionization lines not seen
in Obs.\ 1 -- Obs.\ 3 are present in Obs.\ 4, including \ion{C}{3}
$\lambda$1125.6~\AA, \ion{Si}{3} $\lambda$1144.3~\AA, the \ion{O}{3}
triplet at $\lambda$1150~\AA -- $\lambda$1154~\AA, and \ion{C}{4}
$\lambda$1169~\AA.

The Obs.\ 4 spectrum is similar to a Hopkins Ultraviolet Telescope
(HUT) spectrum of U~Gem acquired 11 days after the end of an outburst
of similar duration \citep{long1993}.  The fluxes in the FUSE spectrum
are about 30\% higher than in the HUT spectrum, which, for example,
has a flux at 1000~\AA\ of \expu{3.5}{-13}{$\flam$}.  The quiescent
HUT spectrum of U~Gem was well fit by WD model atmosphere spectra.
The similarity between the FUSE and HUT spectra indicates that the
Obs.\ 4 FUSE spectrum is also dominated by the WD, a point that will
be addressed in detail in \S\ref{sec_obs4}.

The Obs.\ 4 continuum fluxes did not vary over the course of the
observation.  The mean flux in the time-averaged spectrum is
\expu{5.05}{-13}{$\flam$} in a line-free region, 1045 -- 1055~\AA.
None of the 300 sec spectra have a mean flux in that region more than
4\% from the time-averaged mean; the standard deviation of the (1045
-- 1055~\AA) means in the individual spectra with respect to the
overall mean is 2.6\%.  An examination of count rates from the raw
data files also shows insignificant continuum variability on 10 sec
time scales in Obs.\ 4.

Line identifications, EWs, and FWHM for the FUV lines intrinsic to
U~Gem are given for the time-averaged Obs.\ 1 and Obs.\ 4 spectra in
Table~\ref{tab_ew} (the EWs and FWHM for Obs.\ 1 are representative of
those found for Obs.\ 2 and Obs.\ 3).  Line identifications,
oscillator strengths, and EWs of the prominent interstellar lines in
the U~Gem spectrum are given in Table~\ref{tab_ism}.  The absorption
line parameters were found by fitting Gaussian profiles to the lines
using the IRAF task \textit{splot}.  The EWs for the interstellar
lines were obtained primarily from the Obs.\ 1 time-averaged spectrum.
A few of the interstellar lines (particularly those in the 1000 --
1050~\AA\ range, where all four optical channels contribute to the
data) were remeasured in individual segment spectra to minimize
uncertainties caused by slight differences in the wavelength solutions
of the individual channels.

\section{The Outburst Plateau: Obs.\ 1 -- Obs.\ 3} \label{sec_obs123}

\subsection{The Accretion Disk Continuum} \label{sec_disk}

In outburst, the UV continua of DN are dominated by emission from
the accretion disk.  The slopes of the UV continua resemble those
expected from steady state disks modelled as
appropriately-weighted sums of blackbodies or stellar spectra.
Since the predicted continuum shapes vary, one can in principle
use such models to determine the mass accretion rate in the inner
portion of the disk.  In the case of U~Gem, \citet{panek1984},
using IUE spectra, estimated a mass accretion rate equal to
$\dot{m}$ = \expu{5}{17}{g \: s^{-1}}, or \expu{8}{-9}{\msun \:
yr^{-1}}.  More recently, \citet{sion1997}, using a GHRS data set
of high spectral resolution but limited wavelength coverage, found
$\dot{m}$ = \expu{2.1}{-9}{\msun \: yr^{-1}}.  Neither of these
observations extended below 1200~\AA, however.  Observations at
the longer UV wavelengths sample the Rayleigh-Jeans side of the
blackbody curve for the typical peak disk temperatures in outburst
($\sim$50,000~K) and, as a result, small changes in spectral slope
result in large changes in the estimated mass accretion rate.

Our FUSE spectra probe the short FUV wavelengths where the shape of
the spectrum is presumably a more sensitive tracer of the mass
accretion rate through the inner disk.  Therefore, we generated disk model
spectra to test whether disk model fits to the FUV spectra of U~Gem in
outburst are consistent with those fit at longer wavelengths.  Because we
have outburst observations from three epochs, we can also test whether
the disk shows changes in its structure on decline from outburst.  We
synthesized the model accretion disk spectra using the model
atmosphere and spectral synthesis codes of Hubeny, TLUSTY and SYNSPEC
\citep{hubeny1988,hubeny1994}. Model spectra of steady-state accretion
disks were constructed from summed, area-weighted, Doppler-broadened
spectra of stellar atmospheres of the appropriate effective
temperature and gravity for each disk annulus (see Long et al.\ 1994b
for a detailed description of this procedure).

For modelling the spectra, we assumed an inclination angle $i =
67\arcdeg$, E(B--V) = 0, and $N_{H}$ = \expu{2.0}{19}{cm^{-2}}.
The neutral hydrogen column density was found by a curve of growth
analysis of the UV interstellar lines (see Appendix~\ref{sec_cog};
the assumed column density affects only the cores of the Lyman
series lines, so our fits are actually insensitive to the chosen
value of $N_{H}$.) The fits have two variables: a mass accretion
rate, $\dot{m}$, and an overall normalization, which is 1 for a
distance of 96 pc.  Since there are many relatively narrow
features in the spectra, originating both in the interstellar
medium and in U~Gem, we adopted a modified $\chi^{2}$ approach.
Specifically, we first attempted to fit all of the data in the
spectrum.  We then discarded for the purpose of fitting all data
points that deviated by more than $10 \: \sigma$ from the model
and refit the data.  This process effectively discards the most
discrepant points, is relatively unbiased, and produces fits that,
to the eye at least, model the portions of the data we expect be
able to model in this manner.  (This approach also has the
advantage that one can easily handle more complex models than the
one being considered here, and that, in our case, the results were
also insensitive to the exact cutoff value for rejecting data
points that was chosen.)

The ``best fit'' disk model to the Obs.\ 1 spectrum is shown in
Figure~\ref{fig_d1_disk}. The model spectrum resembles the shape
of the observed spectrum, including the broad \ion{H}{1}
absorption at 1025~\AA.  The main model failure is at the shortest
wavelengths, where it under-predicts the flux. The mass accretion
rate in the model is \expu{6.9}{-9}{\msun \: yr^{-1}} and the
normalization is 0.83. This mass accretion rate is similar to that
determined from the data longward of Ly$\alpha$.  Although the
normalization formally suggests a distance about 10\% greater than
the astrometric distance, this is actually an indication that the
spectral shape and the normalization are in pretty good agreement.
This was not guaranteed. Model fits to spectra taken with HUT have
shown that when the observed FUV peak and downturn in flux to the
Lyman limit are included, steady-state disk models are often
unsuccessful at replicating both the shape and the fluxes of disk
spectra \citep{long1994b,knigge1997}.

Similar fits to the data from Obs.\ 2 and Obs.\ 3 produced similar
results.  This was expected from the fact that the spectra themselves
are nearly identical, and the small changes in the model parameters
are not in our view indicative of a change in the mass accretion rate.
The best fit model for Obs.\ 2 had a nearly identical $\dot{m}$ of
\expu{7.05}{-9}{\msun \: yr^{-1}} and a normalization of 0.77, while
Obs.\ 3 yielded $\dot{m}$ of \expu{5.11}{-9}{\msun \: yr^{-1}} and a
normalization of 0.95.

As noted, the primary problem with all of the models occurs shortward
of 950~\AA.  It is not clear whether this is a problem in the disk
models or whether a second source of emission is indicated.  The disk
models being used are created by summing spectra for atmospheres of
stars, rather than summing the specific intensities for the
inclination angle of U~Gem.  In addition, the effective gravity varies
with height above the disk.  It is possible that spectral syntheses
using somewhat more realistic models could resolve this discrepancy,
but at present such models do not exist . We also note that we have
not allowed for a boundary layer, and in U~Gem we know, based on EUVE
observations \citep{long1996}, that the boundary layer is luminous.
\cite{long1996} estimated the color temperature of the boundary layer
to be 138,000~K, with a size comparable to the WD radius near the
beginning of the outburst. A 138,000~K, optically thick boundary layer
with a size of $4\times10^{8}$~cm \citep{long1996} would contribute
$\sim$25\% of the observed flux in Obs.\ 1 at 915~\AA\ but only
$\sim$13\% at 1180~\AA.  Re-radiation of boundary layer emission from
the disk could also increase the effect of the boundary layer at the
shortest wavelengths.

\subsection{Orbital Variability in the Absorption Line Spectrum} \label{sec_lines}

Obs.\ 1 -- Obs.\ 3 indicate that U~Gem has a rich absorption line
spectrum while on outburst plateau, with the spectra showing
transitions of \ion{H}{1}, \ion{He}{2} and two- to five-times ionized
transitions of C, N, O, P, S and Si.  The absorption lines are
relatively narrow, with FWHM ranging from 250 -- 850~km~s$^{-1}$.
(The FWHM given for Obs.\ 1 in Table~\ref{tab_ew} are for the
time-averaged spectrum and are therefore affected by orbital smearing
of the line profiles.  Nevertheless, the FWHM given are fairly
representative of the FWHM in the individual spectra, which typically
have line widths within 100~km~s$^{-1}$ of the values in
Table~\ref{tab_ew}.)

With the exception of \ion{O}{6} (see \S\ref{sec_o6}), the lines are
smooth, roughly Gaussian, and show little or no structure.  There are
no emission features or P Cygni line profiles in any of the spectra,
except for \ion{O}{6} $\lambda$1038~\AA, which shows apparent weak
emission to the red of the absorption line.  The line ratios in the
\ion{S}{6} $\lambda\lambda$933,945~\AA\ and \ion{O}{6}
$\lambda\lambda$1032,1038 doublets are 1:1, indicating that these
lines are optically thick.  Neither these lines nor any of the
absorption lines in the spectra become dark (i.e., drop to zero flux)
in their line cores, suggesting that the line absorption region does
not completely cover the FUV continuum source (assuming that the
source function of the absorbing material is negligible compared to
that of the background continuum source).

\begin{figure*}
\plotone{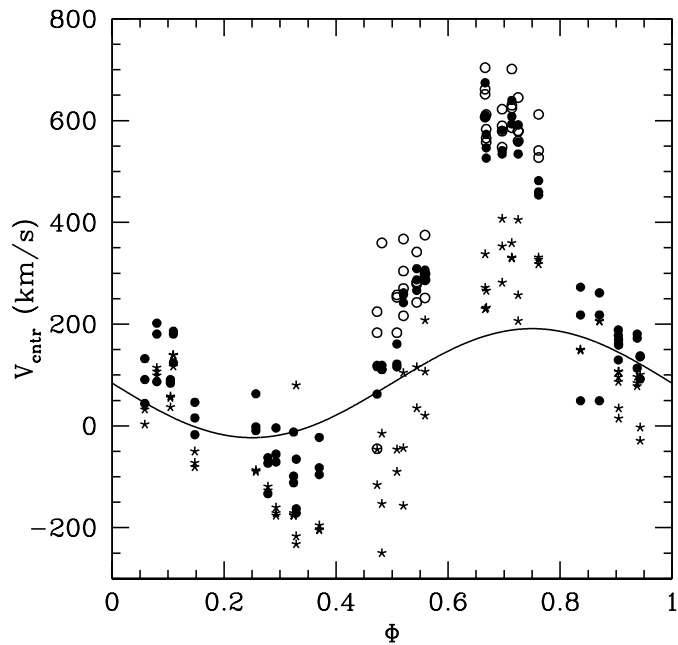} \figcaption[f7.eps]{Velocity offsets from rest
vs. orbital phase for the centers of several absorption lines in the
Obs.\ 1 -- Obs.\ 3 spectra.  The filled circles are the velocities of
three strong metal lines observed at all orbital phases:
\protect\ion{S}{4} $\lambda$1073~\protect\AA, \protect\ion{P}{5}
$\lambda$1118~\protect\AA\ and \protect\ion{Si}{4}
$\lambda$1122~\protect\AA.  The open circles show three of the weaker
metal lines present only at orbital phases 0.5 -- 0.8:
\protect\ion{N}{3} $\lambda$980~\protect\AA, \protect\ion{S}{3}
$\lambda$1021~\protect\AA\ and \protect\ion{C}{3}
$\lambda$1175~\protect\AA.  The asterisks show the velocities of
\protect\ion{S}{6} $\lambda$944~\protect\AA\ and \protect\ion{O}{6}
$\lambda\lambda$1032,1038~\protect\AA.  The solid line is the orbital
motion of the WD assuming a binary recessional velocity of
84~km~s$^{-1}$ (Wade 1981).  The true WD radial velocity curve also
has a gravitational redshift, which is not shown here. Orbital phase 0
is defined as inferior conjunction of the mass donor star.  The binary
ephemeris was taken from Marsh et al.\ 1990. \label{fig_velo}}
\end{figure*}

The absorption lines are at low velocity: in none of the spectra are
the line centers more than 700~km~s$^{-1}$ from their rest positions.
The absorption lines shift in velocity over the binary orbit.  The
shift is in phase with the motion of the WD; Figure~\ref{fig_velo}
shows the velocities of the line centers versus orbital phase for
several absorption lines.  The amplitude of motion of the WD, shown in
Figure~\ref{fig_velo} and measured from narrow WD absorption lines in
GHRS quiescent spectra of U~Gem, is K$_{1}$ = 107.1~km~s$^{-1}$
\citep{long1999}.  While the absorption line centers show much the
same range of motion as the WD at some orbital phases, many of the
lines are blueshifted in excess of the WD motion around phase 0.25,
and all of the lines are redshifted in excess of the WD motion around
phase 0.75.  The peak redshift exceeds the WD velocity by
$>$400~km~s$^{-1}$.  \ion{S}{6} and \ion{O}{6} show a different range
of motion in their line centers from the other metal lines: they are
not as redshifted around orbital phase 0.75 but are more blueshifted
($\sim$150~km~s$^{-1}$) around phase 0.25.  Because the velocities of
their line centers are obscured by interstellar \ion{H}{1} absorption,
\ion{H}{1} lines are not shown on Figure~\ref{fig_velo}, but at those
orbital phases where their minima are well separated from the
interstellar lines, the \ion{H}{1} velocities are consistent with
those of the lower ionization metal lines.

\begin{figure*}
\psfig{file=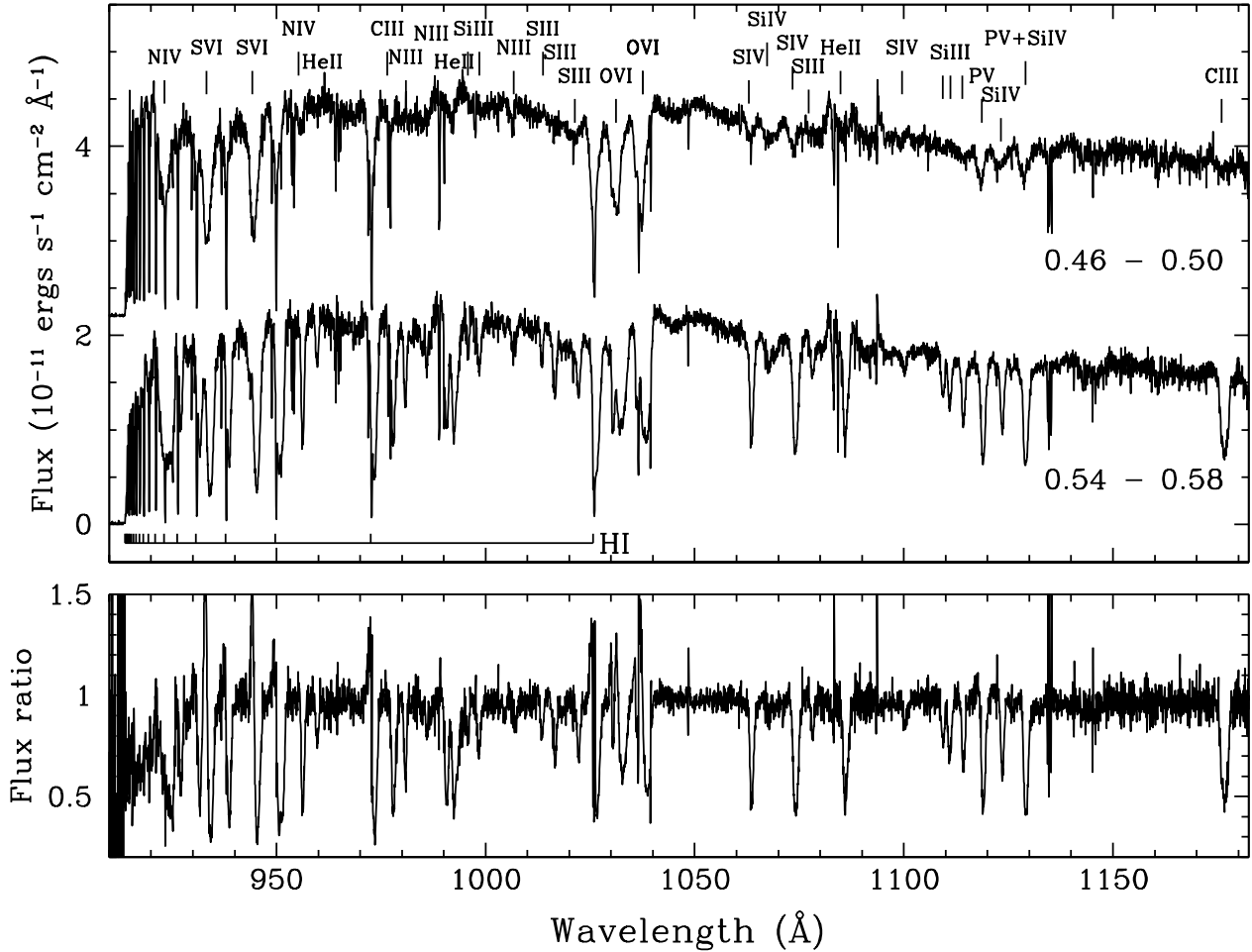,height=5in,angle=-90} \figcaption[f8.eps]{The
outburst plateau spectrum of U~Gem at two orbital phases and a ratio
of the same.  The top panel shows two Obs.\ 1 spectra: the upper
spectrum was obtained over orbital phases 0.46 -- 0.50 and has been
shifted upward by
$2.2\times10^{-11}$~ergs~cm$^{-2}$~s$^{-1}$~\protect\AA$^{-1}$; the lower
spectrum was obtained over orbital phases 0.54 -- 0.58 and is shown in
true fluxes.  The orbital phases of each spectrum are indicated in the
figure.  Prominent absorption lines intrinsic to U~Gem are also
labelled.  The lower panel shows the ratio of the lower to the upper
spectra.
\label{fig_var}}
\end{figure*}

From Obs.\ 1 -- Obs.\ 3, the absorption lines show little day-to-day
change in their properties.  Specifically, spectra acquired at the
same orbital phase but in different observations are comparable in
their absorption line shapes, velocities and EWs.  Variability in the
absorption lines on shorter timescales---namely, correlated with the
orbital phase---is seen in all three observations, however.  In
particular, those spectra acquired roughly between orbital phases 0.5
and 0.8 show an increase in the strength and richness of the
absorption line spectrum compared to spectra acquired over the rest of
the orbit.  This is illustrated in Figure~\ref{fig_var} for Obs.\ 1.
The upper panel of this figure shows two spectra, one from orbital
phases 0.46 -- 0.50 and the other from 0.54 -- 0.58.  The lower panel
shows the ratio of the latter to the former spectrum, and demonstrates
that nearly all of the lines are affected but none of the continuum is
affected, as if an additional absorbing layer had been placed in front
of the UV source.  The same orbital behavior is seen in Obs.\ 2 and
Obs.\ 3.

The change in the absorption lines manifests itself as an increase in
the central depth of each line with little or no increase in the line
FWHM.  The increase in the absorption line depth is seen in every
absorption line in the spectrum, with the exception of the
interstellar lines and the \ion{O}{6} doublet.  In addition, a number
of lines that are not seen or are only weakly present in the spectra
at other orbital phases appear in phases 0.5 -- 0.8.  These include
transitions of \ion{He}{2}, \ion{C}{3}, \ion{N}{3}, \ion{S}{3} and
\ion{Si}{3}.  The increase in the line depth is not accompanied by any
change in the FUV continuum flux, although as mentioned previously,
the flux $<$920~\AA\ drops, which we believe to be due to increased
\ion{H}{1} line absorption near the Lyman limit.  The increase in the
strength and richness of the absorption spectrum, unaccompanied by any
change in the continuum flux, suggests that the absorbing source has
no continuum opacity.  We also know from their doublet ratios that at
least some of the absorption lines are optically thick but not dark in
their line cores. If the absorbing source is a pure absorber --- i.e.,
has a zero source function --- than the increase in line absorption at
these phases indicates an increase in covering factor of the continuum
(the inner disk).

None of Obs.\ 1 -- Obs.\ 3 had full orbital phase coverage and the
orbital phases observed varied, with incomplete overlap between
observations.  As a result, we cannot precisely determine the orbital
range over which the increase in the line absorption is seen.  For the
data in hand, however, the excess line absorption is only seen in
spectra obtained after orbital phase 0.53 and before phase 0.79.

\subsection{The \ion{O}{6} Doublet} \label{sec_o6}

\begin{figure*}
\psfig{file=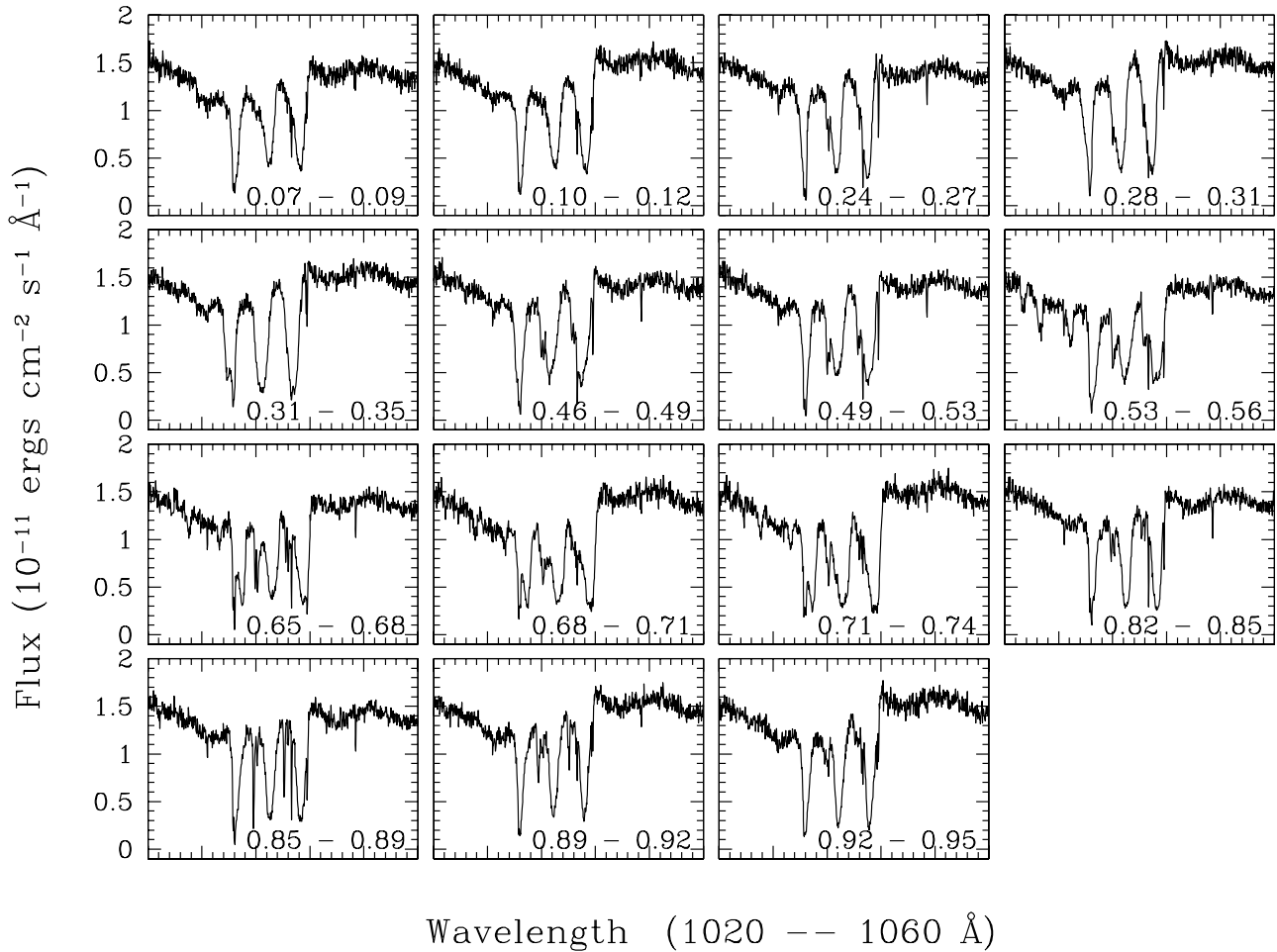,height=5in,angle=-90}
\figcaption[f9.eps]{Spectra of the \protect\ion{O}{6}
$\lambda\lambda$1032,1038~\protect\AA\ doublet in Obs.\ 3.  All of the spectra
are plotted on the same wavelength and flux scales. The orbital phases
covered by each spectrum are indicated. \label{fig_o6}}
\end{figure*}

The behavior of the \ion{O}{6} $\lambda\lambda$1032,1038 doublet is
different from the other absorption lines in the FUV outburst spectra
of U~Gem.  First, the \ion{O}{6} lines are the only lines to show
structure in their profiles.  The line shapes vary dramatically from
spectrum to spectrum; Figure~\ref{fig_o6} shows, for example, the
\ion{O}{6} spectra from Obs.\ 3.  In many spectra, the \ion{O}{6}
lines develop what appears to be narrow ($\sim$100~km~s$^{-1}$),
blueshifted ($-500$~km~s$^{-1}$) dips superimposed on the broader
absorption.  Obs.\ 1 shows such dips in 4 of its 5 spectra.  Eleven of
the 15 spectra in Obs.\ 3 show dips, but oddly, none of the Obs.\ 2
spectra show dips --- the \ion{O}{6} line profiles are smooth
throughout Obs.\ 2.  Although the velocities of the dips can shift by
up to 50~km~s$^{-1}$ from spectrum to spectrum, the shifts in velocity
are not correlated with the orbital motion of the binary.  Many of the
spectra show two dips in each \ion{O}{6} line.  \ion{O}{6}
$\lambda$1038~\AA\ is coincident with two interstellar lines, but
leaving aside distortions in its line shape caused by the interstellar
lines, both \ion{O}{6} $\lambda$1032~\AA\ and $\lambda$1038~\AA\ have
the same profile in any given spectrum.

Second, as mentioned previously, \ion{O}{6} 1038~\AA\ is the only line
in the spectrum that shows any sign of an emission feature, in the
form of a weak bump to the red of the absorption component of the
line.  The strength of the emission feature can vary from spectrum to
spectrum: contrast, for example, the size of the feature in the phase
0.10 -- 0.12 spectrum versus that of phase 0.53 -- 0.56 in
Figure~\ref{fig_o6}. There is no correlation in the strength of the
red emission feature with orbital phase (although it is weakest in the
two Obs.\ 1 and Obs.\ 3 spectra acquired at orbital phase 0.5), nor is
its strength correlated with the shape of the absorption component of
the line.

Finally, while all of the other lines in the Obs.\ 1 -- Obs.\ 3
spectra show an increase in their central absorption depths over
orbital phases 0.53 -- 0.79, the \ion{O}{6} lines do not become deeper
at these phases.  They do become broader in FWHM and FWZI (on their
red wings) at these phases, however.

\subsection{Line Variability in Archival HST Spectra of
U~Gem in Outburst} \label{sec_hst}

U~Gem was observed at high spectral resolution over a limited
wavelength range with the GHRS on HST during two outbursts in 1995
April and September \citep{sion1997}.  The first spectrum, obtained
when U~Gem was on early decline from outburst, is essentially
featureless. The second spectrum, obtained at the peak of the
subsequent outburst, shows narrow (FWHM = 460 -- 640~km~s$^{-1}$), low
velocity \ion{N}{5} $\lambda\lambda$1238,1242~\AA\ and \ion{He}{2}
$\lambda$1641~\AA\ absorption lines superimposed on broad (FWZI
$\sim$2900~km~s$^{-1}$) emission features.  Although the HST spectra
show emission components that are not seen in the FUSE spectral lines,
the shapes and widths of the HST absorption features are consistent
with those of the absorption lines in Obs.\ 1 -- Obs.\ 3.

We retrieved the GHRS data set from the HST archive to see whether the
orbital phase-dependent absorption line variations seen in the FUSE
spectra are also present in the HST spectra.  The HST observations of
\ion{He}{2} $\lambda$1641~\AA\ cover the orbital phases 0.55 -- 0.70 and
0.93 -- 0.96 and therefore include the orbital phases at which the
increased line absorption is seen in the FUSE spectra. We divided
the HST observations into six spectra of 465 sec integration time each
to replicate the time resolution of our Obs.\ 1 -- Obs.\ 3 spectra.

\begin{figure*}
\plotone{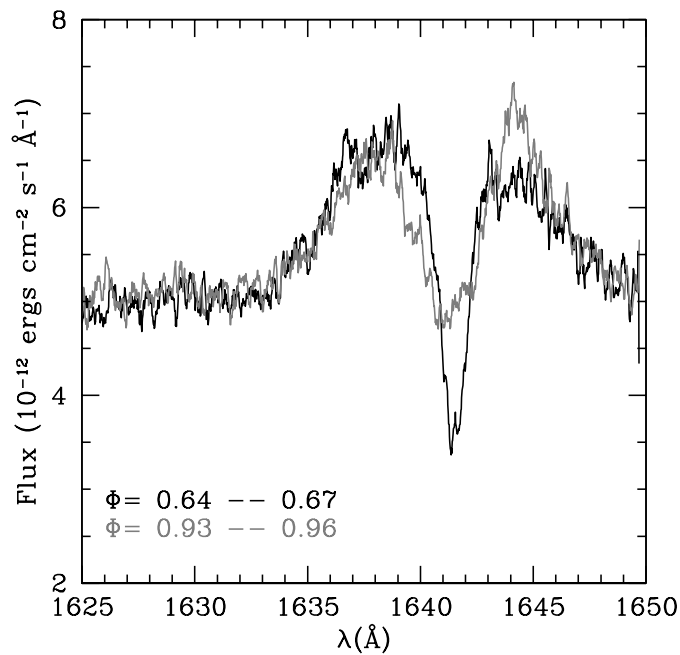}
\figcaption[f10.eps]{HST/GHRS spectrum of \protect\ion{He}{2}
$\lambda$1641~\protect\AA\ in U~Gem at outburst peak, shown at two different
orbital phases.  The spectrum over phases 0.64 -- 0.67 is shown in
black and the 0.93 -- 0.96 spectrum is shown in gray. \label{fig_hst}}
\end{figure*}

The spectra of \ion{He}{2} $\lambda$1641~\AA\ at orbital phases 0.64
-- 0.67 and 0.93 -- 0.96 are shown in Figure~\ref{fig_hst}.  Like the
absorption lines in the FUSE Obs.\ 1 -- Obs.\ 3 spectra, the
absorption core of the \ion{He}{2} feature is deeper at mean orbital
phase 0.655 than at mean phase 0.945.  The \ion{He}{2} absorption line
is relatively shallow in the first spectrum ($\phi_{mean}$ = 0.56),
but it deepens steadily in subsequent spectra, reaching its deepest
level in the $\phi_{mean}$ = 0.655 spectrum shown.  Between
$\phi_{mean}$ = 0.56 to $\phi_{mean}$ = 0.655, the absorption line
shifts to the red by 260~km~s$^{-1}$, a shift well in excess of the
WD's radial orbital velocity motion.  The penultimate spectrum
($\phi_{mean}$ = 0.68) also shows a relatively deep absorption line,
but by $\phi_{mean}$ = 0.945 the absorption component of \ion{He}{2}
$\lambda$1641~\AA\ has reverted to the shallow depth seen at
$\phi_{mean}$ = 0.56.  The increase in the central depth and velocity
of the \ion{He}{2} $\lambda1641$~\AA\ absorption line during orbital
phases 0.58 -- 0.70 shows that the orbital variability in line
absorption seen in our FUSE spectra is not unique to the 2000 March
outburst of U~Gem: the 1995 September outburst observed by HST shows a
similar variability.

We also examined the \ion{N}{5} $\lambda\lambda$1238,1242~\AA\ GHRS
spectra, which were acquired over orbital phases 0.79 -- 0.94.  These
spectra show no variations in the absorption or emission components of
the doublet beyond low-level secular variability.  \ion{N}{5} was not
observed at the orbital phases at which absorption line depth changes
are seen in the FUSE spectra, however. We also note that while the HST
spectra show no \ion{Si}{3} $\lambda\lambda$1293--1303~\AA\ features,
these observations were acquired between phases 0.1 -- 0.3, when
\ion{Si}{3} absorption was also absent from the FUSE spectra; thus,
their absence from the HST spectra may result from the orbital phase
of observation, rather than a true absence of \ion{Si}{3} during any
portion of that outburst.

One interesting feature of the \ion{He}{2} $\lambda1641$~\AA\ orbital
variability is that the absorption line core terminates roughly at the
level of the continuum in phases 0.55 -- 0.58 and 0.90 -- 0.92, but
drops well below the continuum in phases 0.58 -- 0.70.  In contrast,
the \ion{N}{5} absorption is much deeper than the continuum throughout
its observation.  Finally, unlike in the FUSE spectra, where most of
the absorption lines become stronger as well as deeper at orbital
phases 0.53 -- 0.79, the \ion{He}{2} $\lambda$1641~\AA\ absorption
component narrows as it deepens, so that the $\phi_{mean}$ = 0.56 and
0.945 spectra actually show the largest EWs and absorption line
fluxes. Whether this change in the FWHM in \ion{He}{2}
$\lambda1641$~\AA\ is secular or related to the increased line depth
is unclear given the limited data set.

\section{The Decline Phase of the Outburst --- The WD Exposed} \label{sec_obs4}

\subsection{One- and Two-Temperature WD Model Fits}

As noted in \S\ref{sec_morph}, the spectra of U~Gem in Obs.\ 4 differ
from those of Obs.\ 1 -- 3, but are similar to a post-outburst HUT
spectrum of U~Gem obtained 11 days after the return to optical
quiescence \citep{long1993}.  The HUT spectrum was well fit in terms
of emission from the metal-enriched photosphere of a WD with an
average temperature of 38,000~K.  The resemblance of the Obs.\ 4
spectra to the HUT spectrum suggests that the WD is the dominant
source in Obs.\ 4.  To test this, we fit model WD spectra to the Obs.\
4 time-averaged spectrum.  The model spectra were created using TLUSTY
and SYNSPEC \citep{hubeny1988,hubeny1994,hubeny1995}.  In this initial
attempt, we generated LTE model atmospheres and from them constructed
synthetic spectra on a fine wavelength scale ($\delta \lambda \leq
0.01$~\AA).  These were subsequently convolved with a Gaussian (FWHM =
0.15~\AA) to replicate the resolution of the Obs.\ 4 spectrum.  The
model spectra covered a range of WD temperatures from 20,000~K $\leq
T_{WD} \leq$ 95,000~K, and gravities from 7.5 $\leq \log g \leq$
9.5. We tested WD rotation rates of 100~km~s$^{-1}$ and
200~km~s$^{-1}$.  \citep{sion1994,sion1998}.  We adopted a neutral
hydrogen column density of $N_{H}$ = \expu{2.0}{19}{cm^{-2}} (see
Appendix~\ref{sec_cog}) and tested reddening values of E(B--V) = 0 and
0.03, according to the reddening law of \citet{seaton1979}.  For the
metal lines, we tested both solar abundances and the abundances from
\citet{long1999}: a C abundance of 0.1 solar, Si and Al abundances of
0.4 solar, and a N abundance of 4 times solar.  The model spectra were
fitted to the data by least-squares minimization. We masked out the
Ly$\beta$ airglow and the \ion{O}{6} absorption lines; the latter are
not expected to originate on the WD unless its temperature is much
higher than anticipated for U~Gem.  Finally, since the model spectra
are at zero velocity, we shifted the Obs.\ 4 spectrum by
$-172$~km~s$^{-1}$ before fitting to remove the recessional velocity
of the WD \citep{long1999}.

\begin{figure*}
\psfig{file=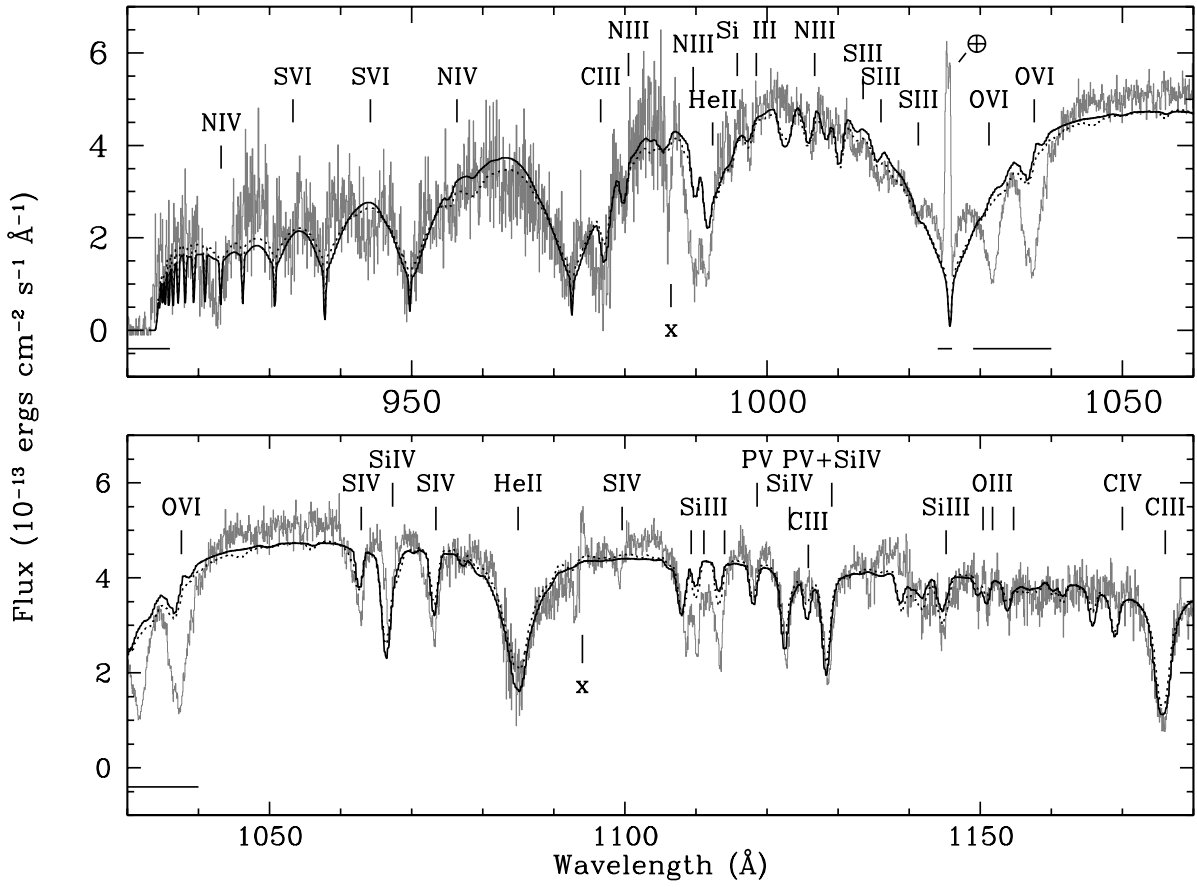,height=5in,angle=-90}
\figcaption[f11.eps]{The Obs.\ 4 spectrum of U~Gem with WD model
spectra superimposed. The solid line shows the best fit single
temperature model, with $T_{WD} = 43,410$~K and $\log g = 8.0$.  The
dotted line shows a two-temperature model with $T_{WD} = 35,000$~K,
$T_{WD,2} = 57,920$~K and $\log g = 8.0$.  The solid bars below the
spectrum indicate the regions that were masked out when the models
were fit to the data. \label{fig_wd}}
\end{figure*}

The best-fit single-temperature WD model for these parameters is shown
as the solid line in Figure~\ref{fig_wd}.  The temperature of the WD
is 43,410~K, the gravity is $\log g = 8.0$, the WD rotation rate is
200~km~s$^{-1}$, and the metal abundances are solar. For the model
shown, the reddening is E(B--V) = 0.03.  The WD model is a good
qualitative (though not statistical: $\chi^{2}_{\nu}$ = 5.67) fit to
the spectrum. The shapes of the continuum and the \ion{H}{1}
absorption lines are reasonably well reproduced by the model.  The
model also reproduces the \ion{He}{2} and metal lines well.  Virtually
all of the lines that are present in the observed spectrum are also
present in the model, except the \ion{N}{4} blend at $\lambda$923~\AA\
and the \ion{S}{6} and \ion{O}{6} doublets.  The strengths of the
lines are mostly well reproduced.  The most glaring exceptions are
\ion{N}{3} $\lambda$990~\AA, \ion{He}{2} $\lambda$992~\AA, and the
\ion{Si}{3} triplet at 1110~\AA, all of which are too weak in the
model to match the observations.  On the other side, the model
predicts strong lines of \ion{N}{3} $\lambda$1002,1003~\AA\ and
\ion{C}{2} $\lambda$1010,1165~\AA\ that are not seen in the data and
overpredicts the strengths of \ion{C}{3} $\lambda$1125.6~\AA\ and
\ion{C}{4} $\lambda$1169~\AA.

The normalization of the model fits is a measure of the solid angle of
the WD and can be combined with the distance to give the WD radius. At
the distance of U~Gem, 96 pc, our normalization gives a WD radius of
$R_{WD} = 4.95\times10^{8}$~cm.  This radius is consistent within
uncertainties in flux calibration with that found by \citet{long1999}:
for the 96~pc distance, their model normalization gives $R_{WD}$ =
$5.5\times10^{8}$~cm.  Assuming a standard WD mass-radius relation
\citep{anderson1988}, our radius indicates a WD mass of 1.08~\msun,
which is close to the mass reported by \citet{sion1998} and
\citet{long1999}, $M_{WD} \sim$1.1~\msun\ (or 1.12~\msun\ for $i =
67\arcdeg$).

Both the temperature and the normalization of the single-temperature
model are largely impervious to variations in the assumed WD rotation
rate, metal abundances, or neutral hydrogen column density: there was
$<$1000~K variation in the WD temperature as these values were varied,
and a $<$5\% change in the normalization (or $\leq$2.5\% change in the
WD radius).  This is not surprising, since the fit is largely set by
the fluxes and the shape of the continuum.  The model parameters are
sensitive to the reddening, however.  For zero reddening, the WD
temperature decreases by a small amount, to 41,750~K, but the
normalization drops by 20\%, or an 11\% decrease in the WD
radius. However, the $\chi^{2}_{\nu}$ only increases from 5.65 to 5.75
and the qualitative fit is essentially unchanged.  Our models are
subject to such uncertainties because the total wavelength coverage of
the FUSE spectrum is small.

Previous studies of the WD in U~Gem \citep{long1993,cheng1997} have
found evidence of a second temperature component, or ``belt'' on the
WD surface.  Given the importance of two-component models to previous
fits to the WD spectrum in U~Gem, we tested the effects of adding a
second temperature component to out WD models.  The best-fit
two-component model, using the same model parameters discussed above
for the single temperature case, is shown as the dotted line in
Figure~\ref{fig_wd}.  It has $\log g = 8.0$, $T_{WD}$ = 35,000~K and
$T_{WD,2}$ = 57,920~K.  A reddening of E(B--V) = 0 gives a slightly
better statistical fit in this case. The ratio of the normalizations
of the two components indicates that the hot component covers some
$\sim$40\% of the WD surface in projected area.  The two-component
model is a statistical improvement in the fit quality: the reduced
$\chi^{2}$ declines from 5.67 to 5.18. This improvement is primarily
manifested as a better fit to the observed spectrum for $<$960~\AA.
At longer wavelengths, the two-component model is very similar to the
single-temperature case. The relative size and temperature of the hot
component is not well constrained by our fits.  By making small
changes to the model parameters, we found that the temperature of the
hot component could vary from 55,000~K to 70,000~K without
significantly changing the fit to the data.  Changes in temperature of
the second component are offset by changes in the size of the hot
region, from a minimum of 13\% to the 40\% noted above.

For the models shown in Figure~\ref{fig_wd}, we assumed solar metal
abundances.  We also tested models in which the metal abundances were
varied.  \citet{sion1998} and \citet{long1999}, modelling separate and
independent observations of U~Gem in quiescence, found C and Si to
have sub-solar abundances and (in the latter study) N to be supersolar,
while the other metals had normal abundances.  We tested the values
found by \citet{long1999}, setting C to 0.1 solar, N to 4 times solar,
and Si to 0.4 solar.  Since these abundances are based on fits to
lines in the HST wavelength range, our FUSE observations should be
well suited to provide an independent check on their values.  For the
elements in question, our models indicate results consistent with
those found by the previous studies.

For carbon, the solar abundance models shown in Figure~\ref{fig_wd}
fit the strong \ion{C}{3} $\lambda$1186~\AA\ and the $\lambda$977~\AA\
lines moderately well. However, most of the other C lines suggest
lower abundances.  Specifically, there are many C lines in the solar
abundance model spectrum that are not present in the data, and other C
lines in the data are strongly overpredicted by a solar abundance
model. For a 0.1 solar C abundance, the fits to the weaker C lines in
the spectrum are vastly improved: lines such as \ion{C}{2}
$\lambda$1066~\AA, \ion{C}{3} $\lambda$1125~\AA\ and \ion{C}{4}
$\lambda$1169~\AA\ are well fit by the 0.1 solar model and the C lines
not present in the data disappear from the models. Since the
\ion{C}{3} $\lambda$1186~\AA\ and the $\lambda$977~\AA\ lines are
saturated, they are not as affected by the abundance. In particularly,
the strongest line, \ion{C}{3} $\lambda$1186~\AA, is still fairly well
fit, but is either slightly too shallow or too narrow, depending on
the WD rotation rate adopted.  The resonance line at $\lambda$977~\AA,
already slightly too weak at solar abundance, becomes even weaker.

\citet{long1999} found nitrogen to be 4 times solar.  This abundance
was based on fits to a single feature, \ion{N}{3}
$\lambda$1184~\AA. There are several \ion{N}{3} and \ion{N}{4} lines
in the FUSE FUV range.  At solar abundance, the resonance line,
\ion{N}{3} $\lambda$990~\AA, is much too weak, as is \ion{N}{4}
$\lambda$955~\AA. At an abundance 4 times solar, both lines are much
better fit; if anything, they remain slightly too weak.  \ion{N}{3}
$\lambda$980~\AA, though not poorly fit at solar abundance, is also
better fit at 4 times solar.  The fits to these lines appear to
confirm the super-solar N abundance found by Long \& Gilliland.  There
are some residual peculiarities in the N line fits, however.  The
\ion{N}{4} blend doesn't appear in any of our models, at any
abundance, despite being a strong feature in the data (and being
present in the model line list).  Also, \ion{N}{3} $\lambda$1002~\AA,
$\lambda$1003~\AA, and $\lambda$1006~\AA\ are too strong in both the
solar and super-solar models.  The overall fit to the N line spectrum,
however, in Obs.\ 4 is improved by assuming a super-solar N abundance.

Our fits are not very sensitive to the abundance of Si.  Most of the
Si lines in the models showed little change in their strengths when
the abundance was changed from solar to 0.4 solar. Within the
uncertainty of our fits, there is no reason to prefer one abundance
over the other.  We note that \ion{Si}{4} $\lambda$1067~\AA\ appears
too strong at solar abundances, but it is blended with the
aforementioned \ion{C}{2} $\lambda$1066~\AA\ line, and a reduction of
the C abundance from solar improves the fit to this feature without
requiring an adjustment to the Si abundance.  Also, the \ion{Si}{3}
triplet, $\lambda$1108~\AA\ -- $\lambda$1113~\AA, is too weak in both
the 0.4 solar and solar models.

The Obs.\ 4 spectrum has WD absorption lines from two elements that
were not observed in the HST spectra: oxygen and phosphorus.  In our
model fits, solar abundances proved good fits to both the \ion{O}{3}
and \ion{P}{5} lines.  Similarly, there are only a few weak sulphur
lines in the HST spectrum, but several in the FUSE spectrum.  A solar
S abundance provides fairly good fits to the \ion{S}{3} and \ion{S}{4}
lines in the FUSE spectrum, although the resonance lines, \ion{S}{4}
$\lambda$1066,1076~\AA, are slightly too weak in the model shown in
Figure~\ref{fig_wd}.

On whole then, our model fits confirm the earlier abundance
peculiarities seen on the WD in U~Gem, both an underabundance of C and
an overabundance of N in the atmosphere. This said we were unable to
accurately determine self-consistent abundance values that fit every,
or almost every, line of a given element. The U Gem spectra are
somewhat cautionary in this regard, suggesting that very careful
modelling is required to accurately assess the errors in derived
abundances. In our case, we have used LTE models; one might hope that
non-LTE modelling and/or very careful attention to the assumed
oscillator strengths could improve the results. We have not pursued
this here, in part due the fact that U Gem may not have been
completely in quiescence at the time of our Obs.\ 4; disk or boundary
layer material associated with the outburst may be affecting the line
strengths. Possible surface temperature variations on the WD
associated with the outburst could also be affecting the
fits. Detailed, non-LTE modelling of a high quality FUV spectrum in
full quiescence is probably a pre-requisite for a more detailed
analysis of this spectrum.  That will be interesting of course, since
the photospheric abundances may change during the inter-outburst
interval due to gravitational settling in the atmosphere if the
accretion rate is sufficiently low or concentrated on a small part of
the WD.

\subsection{Non-WD Features in the Obs.\ 4 Spectrum}

The Obs.\ 4 spectrum shows transitions of \ion{O}{6} and \ion{S}{6}.
The \ion{S}{6} lines are weaker than in the Obs.\ 1 -- Obs.\ 3 spectra
(see Table~\ref{tab_ew}).  The \ion{O}{6} lines have similar EWs to
those of the plateau spectra, but the lines are smooth and do not show
the variable line profiles seen in the plateau observations.  The
\ion{S}{6} and \ion{O}{6} lines are also present in quiescent spectra
of U~Gem \citep{long1993}. The lines do not originate on the WD, which
is too cool to produce such high ionization potential transitions.
The doublet ratio of \ion{O}{6} indicates that the lines are optically
thick.  The doublet ratio of \ion{S}{6} appears to be intermediate
between 2:1 and 1:1, suggesting the \ion{S}{6} doublet is marginally
optically thick.  We fit simple absorption models to the doublet
lines, assuming uniform absorbing slabs of zero source function.  The
resulting column depths of the lines are $\log N_{OVI}$ = 16.0 and
$\log N_{SVI}$ = 15.1. The lines are not dark in their centers, so the
covering fraction of the absorbing slabs is $<$1: 0.65 for \ion{O}{6}
and 0.35 for \ion{S}{6}.\footnote{The \ion{N}{5} doublet at
$\lambda\lambda$1238,1242~\AA\ is also seen in quiescent UV spectra of
U~Gem and is not believed to originate on the WD. The doublet ratio
indicates that the \ion{N}{5} lines are also optically thick.  Using
the GHRS observations of U~Gem (Long \& Gilliland 1999), we determined
the column depth of the \ion{N}{5} absorption, and obtained $\log
N_{NV}$ = 15.3, with a covering fraction of 0.65.}

We also checked the Obs.\ 4 data to see if the increase in line
absorption during orbital phases 0.53 -- 0.79 seen in Obs.\ 1 -- Obs.\
3 is also seen in Obs.\ 4.  To do so, we averaged all Obs.\ 4 data
acquired during phases 0.53 -- 0.79 and all data acquired at other
orbital phases, and examined a difference spectrum of the two
averages.  The difference showed no change in the Obs.\ 4 spectrum
between the two parts of the orbit.  There was no sign of increased
absorption underlying the WD features, nor was there any change in the
shapes or strengths of the \ion{O}{6} and \ion{S}{6} lines, indicating
that the orbital phase-dependent variations in line absorption seen on
outburst plateau is not observed in late outburst decline.

\section{Discussion} \label{sec_disc}

\subsection{The Outburst Spectra -- Disk Plus Stream Overflow}

During the peak and plateau phase of the outburst, the shapes of the
FUSE spectra were very similar and the flux declined by only
$\sim$22\% overall.  Based on our accretion disk models of U~Gem,
$\sim$90\% of the flux at 1000~\AA\ arises from disk radii
$<$1$\times10^{10}$~cm, or $<$20 WD radii.  Thus, the similarity of
the spectra from the three plateau observations is strong evidence
that the structure of the inner accretion disk was relatively stable
and that the effective size of the emitting region did not decline by
more than about 20\% -- 25\%.  Indeed, the flux decline we observed
was similar to that observed by EUVE during a similar outburst
\citep{long1996}.  In both outbursts, the UV flux dropped earlier than
the optical flux, suggesting that the declines were inside-out.

Our disk model fits to the FUV spectra of U~Gem on the outburst
plateau provide qualitatively reasonable determinations of
\expu{7}{-9}{\msun \: yr^{-1}} for the mass accretion rate.  The
accretion disk models successfully reproduce the overall shape of much
of the FUV continuum, although the model under-predicts the observed
continuum $<$960~\AA.  Given our elementary understanding of accretion
disk atmospheres, it is not clear whether the latter discrepancy is
due to a limitation in the disk models or to the addition of a second
continuum component, such as the boundary layer, to the light curve at
the shortest wavelengths. Interestingly, there is reasonably good
agreement between the accretion rate we derive and the rates that have
been derived at somewhat longer wavelengths using IUE and HST
\citep{panek1984,sion1997}.  Our own view, based on studies of other
dwarf novae, is that this is somewhat fortuitous, mainly because
studies of other DN and novalike CVs have shown that the spectra of
many systems with high-$\dot{m}$ depart from the predictions of
synthetic spectra created from summed stellar spectra
\citep{wade1984,long1991,long1994b,knigge1997}.

UV spectra of DN in outburst in the range accessible to IUE and HST
usually contain features due to resonance lines, mainly of \ion{C}{4},
\ion{Si}{4} and \ion{N}{5}. These lines are usually characterized by
larger velocity widths, blue-shifted absorption centroids, and, in
lower inclination systems and especially for \ion{C}{4}, red emission
wings. Such lines are interpreted as resonant scattering in a wind
driven from the inner portions of the disk. This interpretation arises
naturally from the line shapes of typical systems, which for
\ion{C}{4} in particular resemble the P~Cygni profiles produced in O
star winds; from the fact that even when P~Cygni emission wings are
absent the lines are broad and blueshifted; from blue-edge velocities
of up to 5000~km~s$^{-1}$, which are similar to the escape velocity
from the WD; and from the persistence and shape of the line emission
during the eclipses of highly inclined systems (see C\'{o}rdova~1995
and Drew~1997 for reviews).  This interpretation has been supported by
detailed comparison of line shapes to those predicted based on
radiative transfer codes
\citep{drew1987,mauche1987,shlosman1993,vitello1993,knigge1997}.

\begin{figure*}
\psfig{file=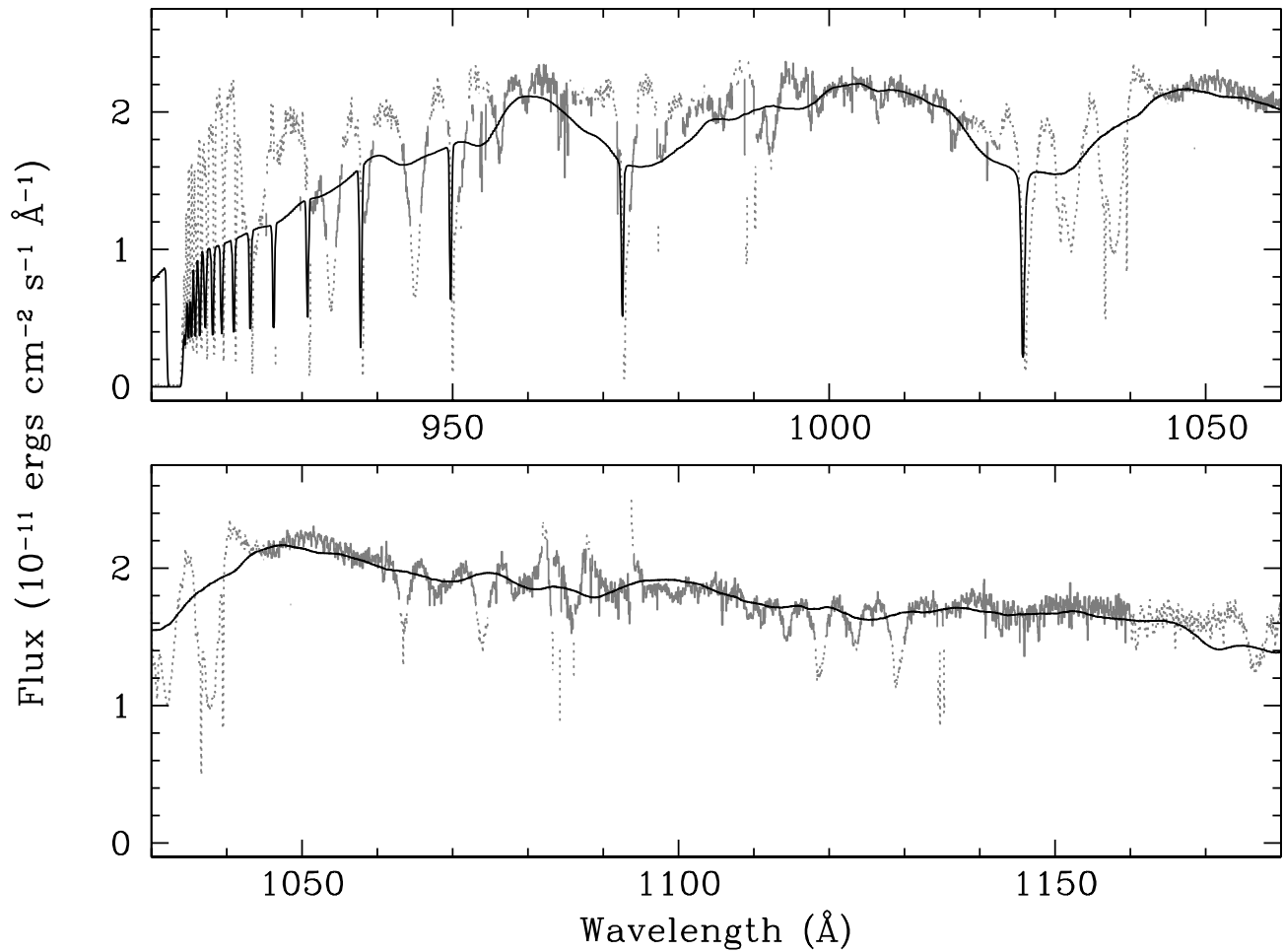,height=5in,angle=-90}
\figcaption[f6.eps]{The Obs.\ 1 spectrum of U~Gem with a model
steady-state accretion disk spectrum superimposed. The parts of the
spectrum plotted as dotted lines are those regions left out of the
final fit in the iterative fitting process. \label{fig_d1_disk}}
\end{figure*}

However, none of these characteristics is seen in the lines in U Gem,
and as a result it is quite clear that a ``standard'' wind does not
create the FUV line structure.  In particular, all of the FUV lines
are relatively narrow with FWHM $\leq 850$~km~s$^{-1}$.  It is also
clear that the lines do not originate directly from the disk
photosphere, since the line features in our steady-state disk models
are broad (see Figure~\ref{fig_d1_disk}).  The reason that
steady-state disk models cannot produce these lines is simple: the
line-of-sight velocities in inner disk where the bulk of the FUV
radiation is produced are simply too high to produce narrow lines.

\begin{figure*}
\psfig{file=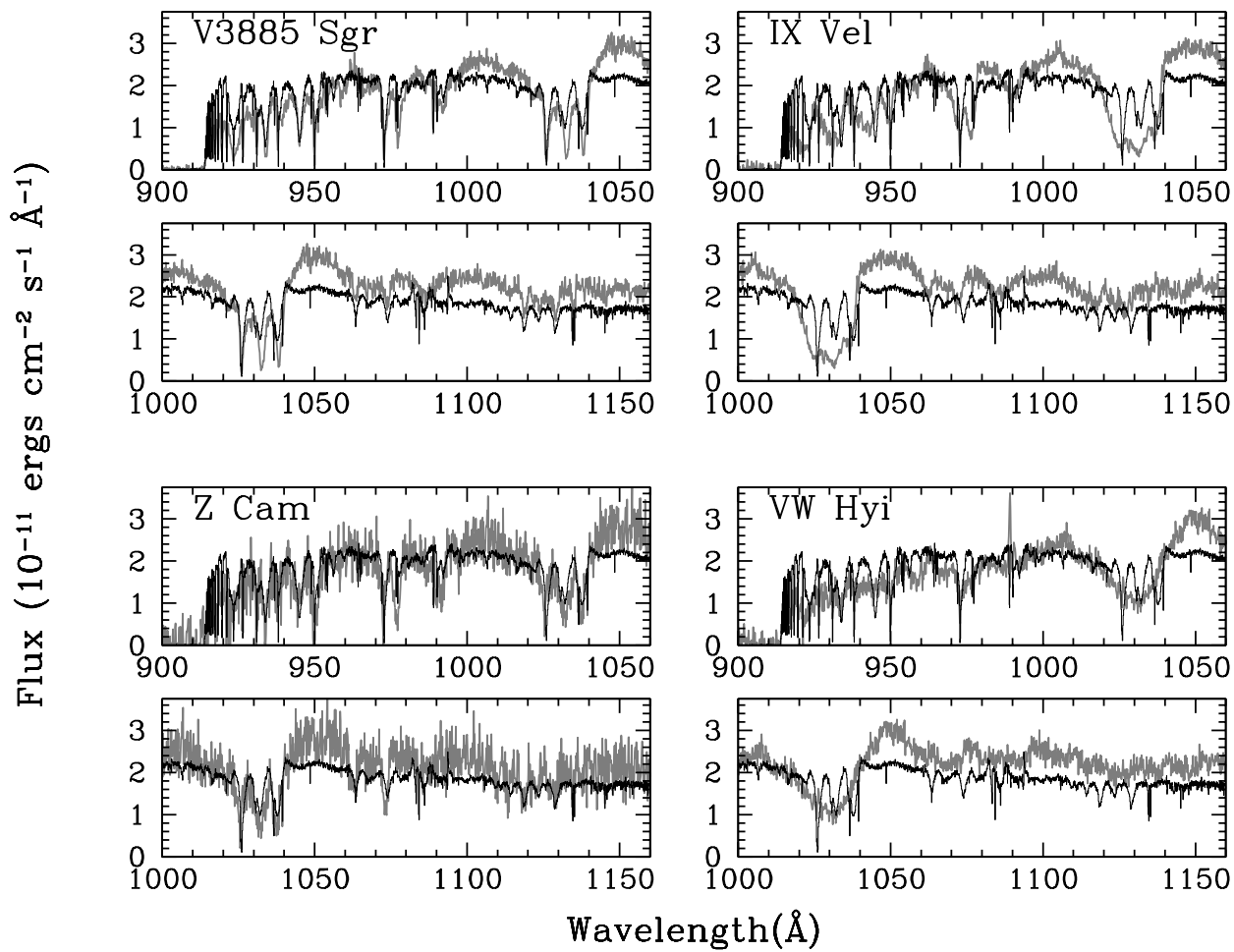,height=5in,angle=-90}
\figcaption[f12.eps]{ORFEUS spectra of the novalike CVs, V3885 Sgr and
IX~Vel, and the DN, Z~Cam (in standstill) and VW~Hyi (in outburst),
compared to the FUSE Obs.\ 1 spectrum of U~Gem. The ORFEUS spectra are
shown in gray; the thin black line in each frame shows the Obs.\ 1
time-averaged spectrum of U~Gem.  The spectra of V3885~Sgr and VW~Hyi
are the exposure-time weighted means of two observations, while the
IX~Vel and Z~Cam spectra are from single observations.  For
comparison, the ORFEUS data have been binned by 10 pixels and scaled
so that the 900 -- 1100~\protect\AA\ mean flux in each spectrum is equivalent
to that of U~Gem in Obs.\ 1. \label{fig_orfeus}}
\end{figure*}

The line spectrum of U Gem in outburst is fairly unusual for a DN in
outburst in the IUE/HST range, but there are relatively few spectra of
CVs covering the FUV range. The largest collection of publicly
available, high spectral resolution FUV spectra of CVs are those
obtained with ORFEUS. (Such FUSE observations are limited at present).
We downloaded spectra obtained by the Berkeley Extreme and Far-UV
Spectrometer (BEFS) on ORFEUS \citep{hurwitz1998,hurwitz1996}. ORFEUS
spectra of the novalike CVs V3885~Sgr and IX~Vel, and of the DN Z~Cam
(in standstill) and VW~Hyi (in outburst) are shown in
Figure~\ref{fig_orfeus}.  To our knowledge, these spectra are
previously unpublished.  For comparison, the fluxes in the ORFEUS
spectra were rescaled so that their mean 900 -- 1100~\AA\ fluxes are
equivalent to the mean flux in the same wavelength range of our Obs.\
1 time-averaged spectrum, which is also shown in each frame of
Figure~\ref{fig_orfeus}.

In fact, we see that the FUV lines and line widths in U~Gem are not
that unusual.  Specifically, the FUV spectra of V3885~Sgr and Z~Cam
are clearly quite similar to the outburst spectrum of U~Gem.  Their
continuum shapes are comparable, and the absorption line transitions
observed and their widths and velocities are virtually identical. This
is true despite the fact that IUE and HUT spectra of V3885~Sgr and
Z~Cam show P Cygni-like \ion{C}{4} profiles (see, e.g., Prinja \&
Rosen 1995). Narrow lines are not universal in the FUV, however. The
IX~Vel spectrum shows much broader absorption features than are seen
in U~Gem, and its metal lines are blueshifted.  The VW~Hyi spectrum
shows strong, broad \ion{O}{6} absorption and little else.  All of the
CVs except VW~Hyi, which lies below the period gap, have similar (4.25
hr -- 6.95 hr) orbital periods to U~Gem and all (except possibly
V3885~Sgr) have inclinations within 10$\arcdeg$ of U~Gem (57$\arcdeg$
-- 60$\arcdeg$; Ritter \& Kolb 1998).  Thus, while high velocity lines
can be seen in some spectra, we conclude that, among FUV spectra of
CVs, U~Gem is not a pathological case.

We are left then with the question: if the FUV lines in U~Gem and
similar systems are created neither by a fast wind nor by the disk
photosphere, where does the absorption arise? We have a number of
clues.  First, the lines encompass a large range in ionization
potential, and many of the lines arise from excited lower states (see
Table~\ref{tab_ew}) with excitation energies of typically 5 --
15~eV. This suggests high plasma density or high radiation density, so
that the lower levels will be excited.  (In LTE, for example, at
25,000~K, the relative population of an excited state with an
excitation energy of 10~eV is still $\sim$1\%.) Indeed, all of the
lines seen in the outburst spectra are also seen in the Obs.\ 4
spectrum that we are confident is dominated by radiation from a WD
photosphere.  Second, the lines have low radial velocities
($\leq$700~km~s$^{-1}$) with respect to the WD, which rules out an
inner disk origin, and none of the absorption lines are dark in their
lines centers, which suggests that the absorbing region only partially
covers the inner accretion disk. Finally, while the absorption lines
are present at all orbital phases, the depths of the low ionization
state lines vary with phase and are deepest between phases 0.53 and
0.79; in the same part of the orbit, most of the lines are redshifted
by $\sim$450~km~s$^{-1}$ with respect to the WD.

The relatively dense, low velocity absorption spectrum suggests an
outer accretion disk chromosphere origin. This is consistent with the
results of \citet{naylor1997}, who found that U~Gem's accretion disk
in outburst shows considerable vertical extent, with the hydrogen
column density along the line of sight reaching up
to $\sim10^{22}$~cm$^{-2}$.  The orbital variability in the FUSE
spectra suggests that the absorption region is also linked to the
source of the X-ray and EUV dips seen in U~Gem.  The increase in line
absorption seen in the FUSE (and archival HST) spectra occurs at
approximately the same orbital phases at which X-ray and EUV
absorption dips occur in U~Gem \citep{mason1988,long1996}. Most of the
observations of dips have been made during outburst, but
\citet{szkody1996} also observed orbital dips in quiescence with
ASCA. Similar phenomena are observed in a group of low mass X-ray
binaries, the so-called X-ray ``dippers''. (See White, Nagase, \&
Parmar 1995 for a review of these systems.)

It is generally believed that the dips in X-ray binaries and analogous
dips in CVs are due to a vertical extension of the accretion disk
created by the interaction between the disk and the mass accretion
stream \citep{white1982}. Two flavors of models of the phenomenon
exist.  In the first, the response of the disk is to bulge --- that
is, to have a different scale heights --- at various azimuths.  For
example, \citet{hirose1991} carried out three-dimensional calculations
that show bulges at positions that will partially occult the disk near
phase 0.8, with less prominent bulges at phase 0.2 and 0.5. Alternate
models were developed by Frank, King \& Lasota (1987) and
\citet{lubow1989}, who proposed that the dips could be accounted for
by stream material skating over the edge of the accretion disk and
piling up at the co-rotation radius. \citet{frank1987} argued that the
interaction between new and old material results in a general
thickening of the disk at the co-rotation radius, resulting in
prominent dips between orbital phases 0.3 and 0.8.  More recent
calculations by \cite{armitage1998} are beginning to synthesize the
bulge and stream extremes. They indicate that the nature of the flow
depends, as one might expect, on the efficiency of cooling of the
stream at the point where the stream encounters the disk. At low
accretion rates, the stream flows more or less ballistically to the
co-rotation radius; at higher rates, the stream is disrupted,
``splashes'', near the disk rim and appears as a bulge with a phase
dependence reminiscent of the predictions of \citet{hirose1991}.

In terms of general character, our plateau absorption spectra appear
consistent with vertically extended absorbing material interacting
with some kind of stream overflow. In fact, in their analysis of the
ASCA observations of U~Gem, \cite{szkody1996} noted that such a
vertically extended stream, in addition to producing X-ray dips,
should also produce very optically thick absorption lines of low
ionization metals with line widths of a few hundred km~s$^{-1}$, much
as we observe with FUSE.  The increase in line absorption and redshift
at orbital phases 0.55 -- 0.71 in the FUSE spectra is, in this
picture, caused by the presence of additional, in-flowing material as
the disk-stream interaction region moves into our line of sight.
Because we observe absorption lines at every orbital phase, however,
the line absorption region cannot be narrowly confined to the
disk-stream interaction region, but rather indicates an
azimuthally-varying disk chromosphere, disturbed by the stream
overflow.

Ideally, one would at this point develop a detailed model of the
ionization structure of and radiation flow through such a stream
interaction region of the disk.  We do not have in hand a detailed
picture of the density structure of this region, nor do we have in
hand the 3-d radiative transfer code that would be required.  We do,
however, possess a 2-d Monte-Carlo code \citep{long1998} that was
simple to modify so that it could handle an accretion disk
chromosphere.  Our goal was limited to determining the location and
velocity of absorbing material that can produce narrow absorption
lines at low velocity; we did not try to find the ionization structure
of the material or fit the full spectrum simultaneously.  We created a
disk chromosphere that rotates at the Keplerian velocity of the
underlying disk, and ran pure absorption (i.e., zero source function)
calculations through that chromosphere, with the inner accretion disk
continuum discussed above as input, to determine the radial extent and
velocity of the region.

We found two main results. First, as expected, the absorbing
material must have a mean velocity with respect to the line of
sight that is nearly zero.  Any outflow component to the gas
resulted in lines that were too heavily blueshifted. Second, we
were able to produce absorption lines similar in velocity, shape,
and depth to those observed when the absorbing gas was restricted
to outer disk radii, $R \sim 1\times10^{10}$~cm --
$3\times10^{10}$~cm ($\sim$20 -- 60~R$_{WD}$).  As many of the
absorption lines are optically thick, the extent of the absorbing
region, i.e. its covering factor, was more important to matching
the absorption lines than the density.  The model chromosphere has
a vertical scale height of 5$\times 10^{9}$~cm, much greater than
the scale height of disk photosphere, and a base density of
$10^{13}$~cm$^{-3}$. At a disk radius of \expu{2}{10}{cm} and an
inclination angle of 67$\arcdeg$, the line of sight to the WD has
a height above the disk of \expu{\sim8}{9}{cm}, so this is
material that is elevated considerably above the disk photosphere.

The parameters of our simple model for the FUV absorbing region
are consistent with the location of the X-ray dip absorbing
material. The vertical scale height of the FUV absorbing material
is $\sim$0.2 -- $0.6\: R$ (where $R$ is the radial distance to the
WD); Long et al.\ 1996 gave a scale height of 0.42$R$ for the
X-ray/EUV absorber. The inner radius of the absorbing region in
our model (10$^{10}$~cm) is also quite close to the projected
circularization radius of the mass stream for U~Gem,
1.1$\times10^{10}$~cm, and could therefore coincide with the
stream overflow radii \citep{frank1987}.  The outer radius in our
model is 3$\times10^{10}$~cm. This is smaller than the outer
radius of the accretion disk in quiescence ($\sim 4\times10^{10}$
cm; Marsh et al.\ 1990) and thus is perhaps less consistent with
an outer disk bulge picture, though a more sophisticated model is
clearly necessary to demonstrate this in earnest.

The \ion{O}{6} absorption lines behave differently from the other
lines in the FUV spectrum.  In particular, they show signatures of a
wind. In Obs.\ 1 -- Obs.\ 3, \ion{O}{6} $\lambda$1038~\AA\ has
apparent weak emission to the red of its blueshifted absorption
component. Another aspect of the \ion{O}{6} doublet that marks it out
as different from the other absorption features is the appearance in
the doublet lines of narrow absorption dips at a blueshift of
$\sim$500 km~s$^{-1}$.  This lends weight to the identification of the
\ion{O}{6} profiles as partially or mainly wind-formed in that similar
blueshifted absorption dips have also been found in very clearly
wind-formed line profiles in HST/STIS spectra of the non-magnetic
novalike variables IX~Vel and V3885~Sgr (Hartley et al, in
preparation; see also Mauche 1991, Prinja \& Rosen 1995).  As here,
the narrow absorption dips seen in IX~Vel and V3885~Sgr show some EW
modulation (e.g. they are seen to disappear over some minutes in
\ion{Si}{4}~$\lambda$1397 in IX~Vel's spectrum), along with only
slight variation of blueshift. The origin of these dips is as yet
quite unclear. We can comment on what they are not: their persistence
at essentially the same blueshift, when present, tends to rule out the
models that are employed to explain the moving ``DACs'' (discrete
absorption components) in O star winds (see e.g. Prinja et al 1992 and
Cranmer \& Owocki 1996).  Instead, we need a model that invokes a
structure in the disk and/or wind that is nearly azimuthally symmetric
and is able to induce a pile-up of outflowing O$^{5+}$ ions at the
observed $\sim$500 km~s$^{-1}$ blueshift.

If a fast wind exists in U~Gem, as the \ion{O}{6} line profiles
suggest, it must be tenuous enough or ionized to the point where it
has little effect on the spectrum in the FUV, except for a weak
signature of outflow in the highest excitation FUV transition.  The
degree of ionization may be the key difference between U~Gem and other
systems, since U~Gem is known to have a luminous boundary layer during
outburst \citep{long1996}. Further evidence of the wind has to be
sought in the EUV domain where, indeed, very highly ionized
transitions are seen in emission that could originate in the wind
\citep{long1996}.

Given the mass accretion rate estimate to hand and a figure for
the white dwarf radius, we may estimate the ratio $\Gamma =
L/L_{{\rm Edd}}$ between the accretion and Eddington luminosities
for U~Gem. Using respectively $\dot{m} = 7\times10^{-9}$
\msun~yr$^{-1}$ (this paper) and R$_{WD} = 4.7\times10^8$ cm
(Long \& Gilliland 1999), we obtain $\Gamma \simeq 0.001$. If the
wind is driven just by radiation pressure, this value of $\Gamma$
implies, at best, a mass loss rate in the region of $10^{-11}$
\msun~yr$^{-1}$ (deduced from the plot of wind mass loss
rate against $\Gamma$ in Drew \& Proga 2000, see also Proga 1999).
By the standards of high-state non-magnetic CV, this is a weak
wind -- mass loss rates are usually estimated to be an order of
magnitude higher (Drew 1997). In the presence of U~Gem's strong
EUV continuum, we should not be surprised such a wind is highly
ionized.  Thus, while in other high state systems, with denser,
less well-ionized winds and, yet, similar mass accretion rates to
U~Gem, radiation pressure as a driver of outflows may be
insufficient to complete account for the flow (e.g., IX~Vel,
Hartley et al., in preparation), our outburst spectra of U Gem are
compatible with mass loss driven mainly by radiation pressure.

\subsection{The Spectrum in Late Outburst Decline -- the White Dwarf}

Obs.\ 4 was obtained two days before the end of the optical outburst.
In the FUV, the accretion disk had already faded, revealing the white
dwarf; a model WD spectrum with T$_{WD} \simeq 43,000$~K provides a
good qualitative fit to the Obs.\ 4 spectrum.  Previous UV
observations of U~Gem obtained early and late in its quiescent
interval have shown that the WD cools between outbursts
\citep{kiplinger1991,long1994a,long1995,sion1998}.  The WD flux drops
30\% (at 1450~\AA) from early quiescence to late quiescence
\citep{long1995}, and models typically indicate a WD temperature of
38,000~K soon after outburst and 30,000~K several hundred days later
\citep{kiplinger1991,long1994a}.  The fluxes observed in Obs.\ 4 are
30\% higher again (at 1000~\AA) than those seen in a HUT spectrum of
U~Gem obtained 11 days after then end of an outburst, indicating that
the WD luminosity in late outburst is larger still than its luminosity
in early quiescence.  Moreover, the drop in flux from late outburst to
early quiescence is comparable to the decline in flux during the
quiescent interval.

In modelling the HUT spectrum of U~Gem, \citet{long1993} found that a
single-temperature WD model provided a poor fit to the spectrum below
970~\AA.  The quality of the fit at the bluest wavelengths was
improved by the addition of a second component on the WD, a 56,600~K
source occupying $\sim$15\% of the WD surface.  More important, the
additional component resolved an inconsistency between the observed
flux decline during quiescence and the decrease in the WD temperatures
of the quiescent models: an 8000~K drop in the temperature of the
entire WD (from 38,000~K to 30,000~K) would cause the observed flux to
drop by over a factor of two, not just 30\% as observed, but the
cooling of 15\% of the WD from 57,000~K to 30,000~K is consistent with
the observed flux decline.  Indeed, virtually all of the model fits to
quiescent observations of U~Gem are consistent with the bulk of the WD
having a constant or near-constant temperature while a fraction of the
WD --- heated by the outburst --- cools during quiescence
\citep{kiplinger1991,long1994a,long1995,cheng1997}.

Our single-temperature WD model indicates a temperature of 43,000~K.
A decline in the bulk WD temperature from 43,000~K in late outburst to
38,000~K in early quiescence implies a decrease in flux at 1050~\AA\
of 35\%, which matches the observed flux decline within the flux
calibration uncertainties of both spectra.  Therefore, a second
temperature component is not required to explain the drop in flux from
that seen in Obs.\ 4 to those of early quiescent observations of
U~Gem.  Previous observations have indicated the likely presence of a
belt on the WD, however, and if a hot belt is a consequence of the
elevated accretion during outburst, we would expect the belt to also
be present during Obs.\ 4. Our two-temperature WD model is, in fact, a
statistical and qualitative improvement over the single temperature
fit to the Obs.\ 4 spectrum.  The two models are quite similar at
longer wavelengths, but the addition of a second component improves
the fit to the spectrum at the shortest wavelengths ($<$960~\AA), as
was also true of fits to the HUT post-outburst spectrum.

Our observations indicate that the WD is more luminous in late
outburst than in early quiescence and that the addition of a second
temperature component on the WD improves the model fit to Obs.\
4. Unfortunately, we cannot use these results to draw firm conclusions
about changes in temperatures across the WD after an outburst, both
because the effects we are trying to measure are at the margin of what
is possible after allowing for errors in the cross-calibration of the
instrumentation involved, and because there is no guarantee that the
WD behaved identically in the various outbursts that have been
observed (see, for example, Sion et al.\ 1998). The question of
whether all or part of the WD cools after outburst bears directly on
models of the cooling source.  If the heating of the WD during
outburst is caused by spin-up of its surface layers, a multi-component
temperature distribution on the WD in the form of a belt is likely, as
the accreted material preferentially boosts kinetic energies near the
WD equator \citep{long1993}. Ongoing accretion through the disk may
also support a WD belt in quiescence \citep{cheng1997}. Other models
--- such as radiative heating of the WD during the outburst
\citep{pringle1988}, compressional heating of the WD layers by the
weight of accreted material \citep{sion1995}, or quiescent accretion
via a disk corona \citep{meyer1994} --- may allow a single-temperature
WD. A series of targeted observations of U~Gem throughout a full
quiescent interval using HST or FUSE will be needed to resolve
uncertainties concerning the response of the bulk of the WD and the WD
belt to the outburst.

The \ion{O}{6} and \ion{S}{6} absorption lines that are present in the
Obs 4. spectrum do not arise naturally in the WD photosphere, unless
the temperature of the WD is much higher than we have estimated. The
smooth \ion{O}{6} line profiles bear no resemblance to the wind-formed
lines seen on outburst plateau. Also interesting is the fact that none
of these lines (indeed, none of the lines in the Obs.\ 4 spectrum)
show variability, orbital or otherwise. The X-ray dips with which the
outburst plateau absorption spectrum is associated persist into
quiescence \citep{szkody1996}, but no associated orbital variation is
seen in Obs.\ 4.  These lines are not unique to late outburst,
however, as \ion{O}{6} was also observed in the HUT quiescent spectra,
and both the HUT and HST quiescent spectra show \ion{N}{5} absorption.
We derived column densities for the lines of $10^{15.3}$, $10^{16.0}$,
and $10^{15.1} \:cm^{-2}$ for \ion{N}{5}, \ion{O}{6}, and \ion{S}{6},
respectively.  The associated H column densities, assuming close to
solar abundances, for all three lines range is about $10^{19.3}
cm^{2}$. (This is close to that the column density we have derived for
the interstellar column density to U Gem, but that is coincidental
since these are not the dominant stages of these ions in the ISM.)

If accretion onto the WD in quiescence occurs through some kind of
coronal, or siphon flow, such as suggested by \citet{meyer1994}, it is
possible that these high excitation absorption lines arise in the
transition region between the hot X-ray emitting gas and the WD
photosphere. If this is the case, it is possible to estimate the
density $n_e$ and the length scale $\delta r$ from the quiescent
accretion luminosity and the column densities.  Specifically, and
concentrating on \ion{O}{6} \[ n_{e}^{2} \epsilon 4 \pi r^{2} \delta r
\sim L_{OVI region} \sim \frac{GM \dot{m}}{r} \frac{\delta r}{r} \]
where $\epsilon$ is the total emissivity in the \ion{O}{6} region, and
$\delta r$ has been left on both sides of the equation for clarity. If
we assume that the 43,000~K WD temperature we derive our models of
Obs.\ 4 represent excess heating from accretion over the 30,000~K
quiescent temperature of the WD, we obtain an excess luminosity of
$5.6\times10^{32}$~ergs~s$^{-1}$ and a reasonable accretion rate
$2.1\times10^{15}$~g~s$^{-1}$. In coronal equilibrium, the relative
abundance of \ion{O}{6} peaks at $\sim3\times10^{5}$~K and the
corresponding emissivity is
$\sim5\times10^{-22}$~ergs~s$^{-1}$~cm$^{3}$ \citep{raymond1976}, and
this suggests a density of $10^{13} \: cm^{-3}$ in the transition
region between the hot plasma and the WD photosphere.  Since we know
the column density ($n_{H} \delta r$) to be $10^{19.3} cm^{2}$, the
thickness of the region is about 20 km.  This is within an order of
magnitude of the scale height for a $3\times10^{5} \: K$ component to
the upper atmosphere of a log g=8 star.

Finally, one of the more interesting recent discoveries from analyses
of HST UV spectra of U~Gem in quiescence has been that sub-solar
carbon abundances and super-solar nitrogen abundances are required in
the photosphere of the WD. (The models also indicate slightly
sub-solar silicon and aluminum abundances; Sion et al.\ 1998, Long \&
Gilliland 1999). Such an abundance pattern is indicative of a history
of CNO processing in the surface WD material. Intriguingly,
\citet{harrison2000} find that the CO absorption lines in their NIR
spectra of U~Gem are also weaker than expected for the mass donor star
spectral type, which may confirm that the material accreted onto the
WD consists of CNO processed material, perhaps deposited on the mass
donor star during an previous nova explosion in the binary
\citep{sion1998}.  In our models, the overall C and N absorption
spectra are consistent with a sub-solar abundance for the former and a
super-solar abundance for the latter as found by fits to the HST
spectra.

\section{Conclusions} \label{sec_conc}

1. We have obtained FUSE FUV spectra of U~Gem in outburst.  The Obs.\
1 -- Obs.\ 3 spectra, obtained when U~Gem was on the outburst plateau,
are characterized by a curved continuum that peaks at 1005~\AA\ and a
host of absorption lines, both interstellar and intrinsic to U~Gem.
The FUV continuum level declined by 8\% -- 10\% from Obs.\ 1 to Obs.\
2 and 12\% -- 14\% between Obs.\ 2 and Obs.\ 3 with no change in the
shape of the spectrum.  The Obs.\ 4 spectrum, obtained on the late
decline from outburst, is 30 times fainter than Obs.\ 3.  It also has
a bent continuum shape that peaks near 1000~\AA\ and an absorption
line spectrum.

2. The continua of the Obs.\ 1 -- Obs.\ 3 spectra are consistent with
steady-state accretion disk model spectra with a mass accretion rate in
the disk of $\dot{m}$ = $7\times10^{-9}$~\msun~yr$^{-1}$ at peak of
outburst.  The accretion disk models do not fit the absorption lines
in the spectra, which are too narrow to originate in the inner,
FUV-emitting accretion disk.

3.  The non-interstellar absorption lines in the Obs.\ 1 -- Obs.\ 3
spectra are narrow (FWHM $\sim$ 500~km~s$^{-1}$) and low-velocity
($\leq$700~km~s$^{-1}$) transitions of \ion{H}{1}, \ion{He}{2} and
two- to five-times ionized transitions of C, N, O, P, S and Si.  With
the possible exception of a red wing on the \ion{O}{6}
$\lambda$1038~\AA\ line, there are no emission features in the
spectra.  Most of the lines (except those of \ion{O}{6}) are smooth
and show little structure.  The \ion{S}{6} and \ion{O}{6} doublets are
optically thick.  At a given orbital phase, absorption line shapes,
EWs and velocities are similar between observations.  The outburst
spectra of U~Gem are also very similar to ORFEUS spectra of the
novalike CV V3885~Sgr and the DN Z~Cam.

4. The absorption lines show orbital variability. Specifically,
spectra acquired at orbital phases $>$0.53 and $<$0.79 show an
increase in the depths of the absorption lines and the appearance of
numerous low ionization lines not seen at other orbital phases.  A
similar increase in the central depth of its absorption line is seen
in HST spectra of \ion{He}{2} $\lambda$1641 obtained during an earlier
outburst.

5. The absorption line variability occurs at the same orbital phases
as X-ray and EUV light curve dips seen in U~Gem in both outburst and
quiescence. Assuming that the increase in line absorption is tied to
the source of the dips suggests that the absorption line spectrum
originates in extended material above the plane of the accretion disk,
with the orbital variability caused by a disk bulge or mass stream
overflow.  Simple models of the FUSE absorption line spectrum indicate
that the FUV absorbing material must be located at large disk radii in
order to achieve the observed low radial velocity.

6. In late outburst decline, Obs.\ 4, the spectrum is dominated by
the white dwarf.  The fluxes in Obs.\ 4 are 30\% higher than in a
HUT spectrum of U~Gem acquired in early quiescence.  The Obs.\ 4
spectrum is qualitatively well fit by a single-temperature WD
model spectrum with $T_{WD} \simeq$ 43,000~K.  The model indicates
a WD radius of $R_{WD}$ = \expu{4.95}{8}{cm}.  The presence of a
second temperature component on the WD improves the quality of the
model fits to the data at $<$960~\AA.  Our fits confirm the
non-normal metal abundances found in previous studies of the WD.
Optically thick \ion{O}{6} absorption lines and weak, marginally
optically thick \ion{S}{6} absorption are present in the Obs.\ 4
spectrum.  These lines may arise in a transition region above the
WD photosphere.

7.  We conducted a curve of growth analysis of the interstellar lines
in the spectrum of U~Gem.  We find a neutral hydrogen column density
on the line of sight to U~Gem of \expu{2}{19}{cm^{-2}}.  The metal
lines are consistent with a broadening parameter of 5.5~km~s$^{-1}$,
while the interstellar Lyman absorption lines are consistent with $b =
11.5$~km~s$^{-1}$.

\acknowledgements{Based on observations made with the NASA-CNES-CSA
Far Ultraviolet Spectroscopic Explorer. FUSE is operated for NASA by
the Johns Hopkins University under NASA contract NAS5-32985. We wish
to thank the AAVSO, and especially Janet Mattei, for notifying us of
the outburst of U~Gem and monitoring the progress of the outburst.
This was crucial to the success of the observations.  This work would
not have been possible without the heroic efforts of the FUSE staff in
scheduling and conducting the observations.  We would also like to
thank Chris Howk, Meena Sahu, and Chris Mauche for useful suggestions
and comments concerning the interstellar line spectrum and curve of
growth analysis. We also gratefully acknowledge the financial support
from NASA through grant NAG5-9283.}

\begin{figure*}
\plotone{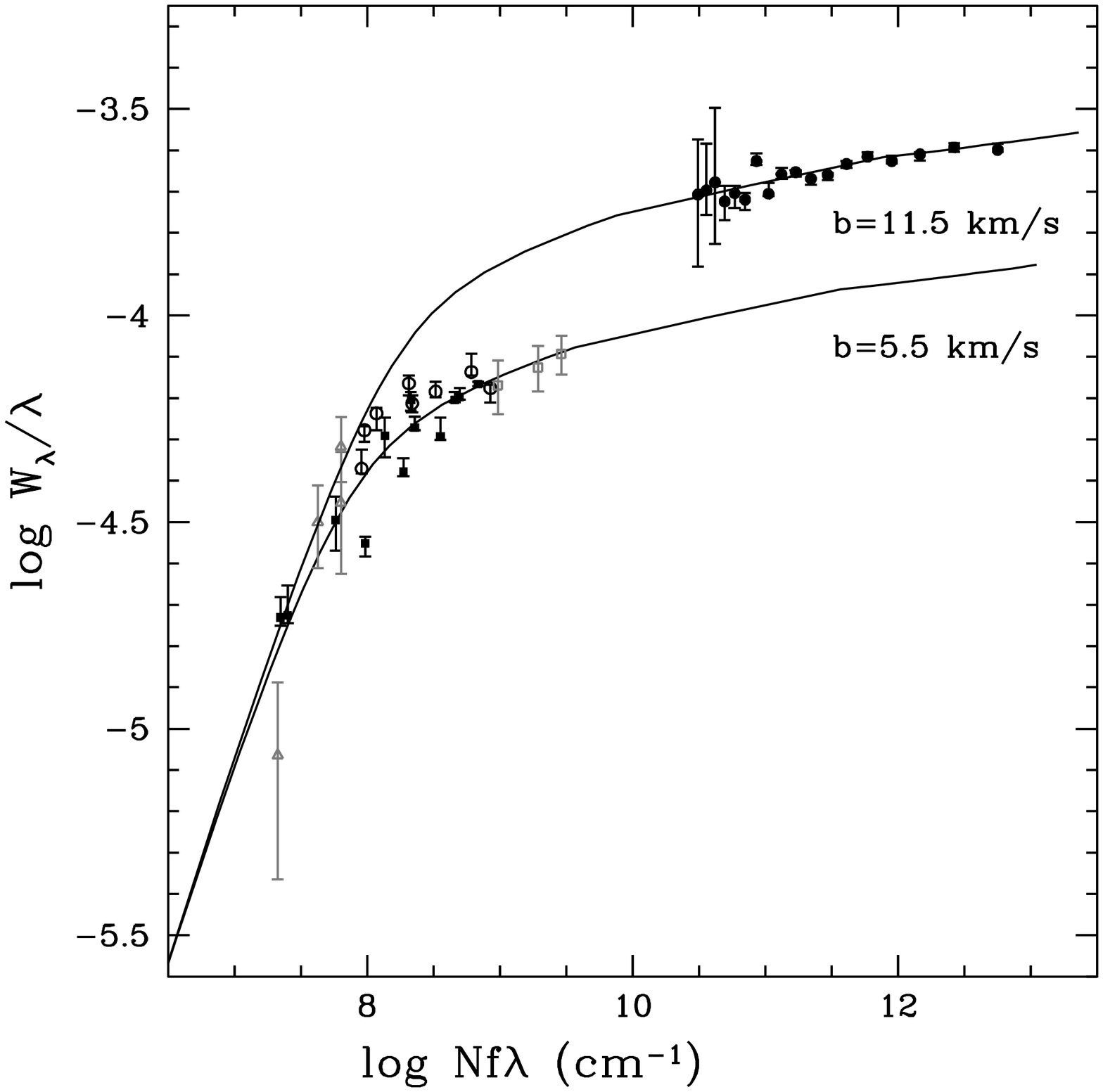}
\figcaption[f13.eps]{The curve of growth for the sight line to U~Gem.
The neutral hydrogen column density for the curves shown is $N_{H} =
2.0\times10^{19}$~cm$^{-2}$.  The broadening parameter for the upper
curve is 11.5~km~s$^{-1}$; $b$ = 5.5~km~s$^{-1}$ for the lower curve.
For the FUSE data points, \protect\ion{H}{1} lines are given as solid
circles, \protect\ion{N}{1} as solid squares, and \protect\ion{O}{1}
as open circles.  The metal abundances have been corrected for their
IS depletion with respect to H using the abundances given by Meyer
(1997,1998).  Also shown are the IUE measurements from Long et al.\
(1996).  The open squares are \protect\ion{N}{1} and the open
triangles are \protect\ion{S}{2}.  Two data points are shown for
\protect\ion{S}{2} $\lambda$1259.520~\AA, which was measured in two
spectral orders. \label{fig_cog}}
\end{figure*}

\appendix

\section{A Curve of Growth Analysis of the Interstellar Lines in the
U~Gem Spectrum} \label{sec_cog}

The interstellar (IS) hydrogen column density on the line of sight to
U~Gem was previously determined by \citet{long1996} from a curve of
growth analysis of the IS metal absorption lines in a high-resolution
IUE spectrum of U~Gem.  Using the method of \citet{mauche1988}, Long
et al.\ determined $N_{H}$ from the growth curve of the metal lines by
making two assumptions: first, that the abundances of nitrogen and
sulpher are undepleted with respect to their solar abundances (and can
therefore be used to determine the hydrogen column density); second,
that a single broadening parameter describes all species.  Using these
assumptions and measurements of \ion{N}{1} and \ion{S}{2} IS lines in
the IUE spectrum, they obtained $N_{H} = 3.1\times10^{19}$~cm$^{-2}$
and $b$ = 5~km~s$^{-1}$.  Because the FUSE wavelength range contains
numerous IS metal transitions as well as IS absorption from the higher
order Lyman series, we resolved to re-analyze the growth curve on the
line of sight to U~Gem.  Although the value of the hydrogen column
density does not affect the models in this manuscript, precise values
of $N_{H}$ are important in the analysis of EUV and soft X-ray
observations of U~Gem.

We used the EWs and oscillator strengths given in Table~\ref{tab_ism}
to construct our growth curve, on which we included unblended lines of
\ion{H}{1}, \ion{N}{1}, and \ion{O}{1}.  We assumed that N and O are
undepleted with respect to their mean gas-phase IS abundances; these
abundances were taken from \citet{meyer1997,meyer1998}.  We
constructed the curve of growth for this sight line, assuming a
single-component Maxwellian distribution (i.e., assuming no damping
wings), using the basic method outlined by \citet{spitzer1978}.  The
curve is shown in Figure~\ref{fig_cog}.  The error bars on the points
represent uncertainties in placement of the continuum only (they are
much larger than the statistical uncertainties for our spectra).  Also
shown in gray are the IUE \ion{N}{1} and \ion{S}{2} points from Long
et al.  The lower growth curve shown is the best fit to the metal
lines: $N_{H} = 2.0\times10^{19}$~cm$^{-2}$ and $b$ = 5.5~km~s$^{-1}$.

Also shown on Figure~\ref{fig_cog} are the \ion{H}{1} absorption lines
from the FUSE spectrum.  The higher-order Lyman series lines can not
be used to directly determine $N_{H}$ because they all land on the
flat part of the curve of growth.  Observations of Ly$\alpha$, which
falls on the square-root part of the growth curve, are necessary to
fix $N_{H}$ from the \ion{H}{1} absorption.  The H lines shown in
Figure~\ref{fig_cog} land above the growth curve defined by the metal
lines.  The upper line in Figure~\ref{fig_cog} shows that if the
broadening parameter, $b$, is increased to 11.5~km~s$^{-1}$, the
growth curve passes directly through the \ion{H}{1} points.  A larger
broadening parameter value is expected for \ion{H}{1}, since Doppler
broadening of the IS line profiles from thermal motions will be larger
for H than for the heavier species.  If $b$ was set solely by Doppler
broadening, however, we would expect it to be $\sqrt{14}$, or 3.74,
times larger for H than for N, while the broadening parameter value
for H is only 2.1 times larger than $b$ for the metal lines in
Figure~\ref{fig_cog}.  This discrepancy can be reconciled if the IS
cloud has an appreciable non-thermal velocity that dominates the
broadening parameter for the metals, while the thermal broadening
dominates for H.  If $b = \sqrt{b_{ntherm}^{2} + b_{therm}^{2}}$ and
$b_{therm,NI} = \sqrt{14} b_{therm,HI}$, then the two $b$ values shown
in Figure~\ref{fig_cog} give $b_{ntherm} = 3.3$~km~s$^{-1}$ and
$b_{therm,HI} = 10.5$~km~s$^{-1}$.  This thermal velocity implies a
cloud temperature of 6600~K.  Note, however, that this calculation
assumes that the effective $b$ values indicated by the curve of growth
are equivalent to the true thermal and non-thermal $b$ values in an IS
cloud.  The \ion{H}{1} are so much stronger than the metal lines that
the presence of low column density clouds in the sight line, for
example, could broaden \ion{H}{1} while not affecting the metal lines
at all.

Note also that the curve of growth analysis is predicated on the
assumption that a single IS cloud determines the IS absorption to
U~Gem.  None of the metal lines (except the uncertain \ion{S}{2}
$\lambda$1250.586~\AA\ line) truly fall on the linear part of the
curve, so the derived $N_{H}$ is subject to uncertainties in the shape
of the growth curve, which would be affected by the presence of
multiple IS clouds.  \citet{dring1997} observed two stars, $\beta$~Gem
($l$ = 192$\arcdeg$, $b$ = +23.4$\arcdeg$, d = 10.3~pc) and
$\sigma$~Gem ($l$ = 191$\arcdeg$, $b$ = +23.3$\arcdeg$, d = 37.5~pc),
on lines of sight similar to that of U~Gem ($l$ = 201$\arcdeg$, $b$ =
+22.7$\arcdeg$).  They found that both these lines of sight pierced
two clouds, for a total $N_{H}$ of $1.75\times10^{18}$~cm$^{-2}$ for
$\beta$~Gem and $1.62\times10^{18}$~cm$^{-2}$ for $\sigma$~Gem.  This
suggests that at least three clouds, perhaps more, are penetrated on
the line of sight to U~Gem, as the column density to U~Gem is an order
of magnitude larger than those to the two stars analyzed by
\citet{dring1997}.  \citet{jenkins1986} concluded, however, that
column densities derived from a curve of growth analysis should only
underestimate $N{H}$ by 20\% at most so long as the distribution of
optical depths and broadening parameters along the sight line is
fairly smooth and highly saturated lines are not present.  The
resolution of our FUSE spectrum ($\sim$20--25~km~s$^{-1}$) is
insufficient to resolve the contributions of individual IS clouds, so
our value of $N_{H}$ is subject to the caveat that a more complex IS
structure than the one assumed in our curve of growth analysis may be
present.  Given that caveat, we find that $N_{H} =
2.0\times10^{19}$~cm$^{-2}$ is the best estimate of the line of sight
column density to U~Gem.

\clearpage
\begin{deluxetable}{ccl}
\tablecaption{Adopted System Parameters\label{tab_par}}
\tablewidth{0pt}
\tablecolumns{3}
\tablehead{ \colhead{Parameter} & \colhead{Value} & \colhead{Reference} }
\startdata
$P_{orb}$ & 0.17690619 d & Marsh et al.\ 1990 \\
$K_{1}$ & $107.1\pm2.1$ km s$^{-1}$ & Long \& Gilliland 1999 \\
$i$ & $67\arcdeg\pm3\arcdeg$ & Long \& Gilliland 1999\\
$M_{WD}$ & $1.14\pm0.07$ \protect\msun & Long \& Gilliland 1999\\
$M_{R}$ & $0.41\pm0.02$ \protect\msun & Long \& Gilliland 1999\\
$d$ & $96.4\pm4.6$ pc & Harrison et al.\ 1999 \\
\enddata
\end{deluxetable}

\clearpage
\begin{deluxetable}{ccccccc}
\tablecaption{Observation Summary\label{tab_obs}}
\tablewidth{0pt}
\tablecolumns{7}
\tablehead{
\colhead{} & \colhead{Date (UT)} & \colhead{Start (UT)} &
\colhead{$\Phi_{Start}$} & \colhead{End (UT)} &
\colhead{$\Phi_{End}$} & \colhead{t$_{obs}$ (sec)} }
\startdata
Obs. 1 & 2000 March 05 & 16:36:37 & 0.46 & 18:43:43 & 0.96 & 2883 \\
Obs. 2 & 2000 March 07 & 10:11:18 & 0.26 & 14:04:01 & 1.17 & 6248 \\
Obs. 3 & 2000 March 09 & 13:50:17 & 0.42 & 21:04:11 & 2.12 & 7868 \\
Obs. 4 & 2000 March 17 & 11:43:20 & 0.14 & 20:34:16 & 2.25 & 12975 \\
\enddata
\tablecomments{Obs.\ 1 consists of five spectra with integration times
of 576 -- 577 sec each.  Obs.\ 2 is divided into nine
spectra of roughly 700 sec exposure time each, and the Obs.\ 3 data
consist of sixteen spectra with exposure times of 423 -- 527 sec each.  The
Obs.\ 4 data were acquired in time-tag mode at high time resolution.
We divided the Obs.\ 4 data into 42 spectra with exposure times of 300
sec (see text).}
\end{deluxetable}

\clearpage
\begin{deluxetable}{cccccc}
\tablecaption{Prominent Lines in the U Gem FUV Spectrum\label{tab_ew}}
\tablewidth{0pt}
\tablecolumns{6}
\tablehead{
\colhead{} & \colhead{} & \multicolumn{2}{c}{Obs.~1} &
\multicolumn{2}{c}{Obs.~4} \\
\colhead{Ion} & \colhead{$\lambda_{lab}$} &
\colhead{EW (\protect\AA)} & \colhead{FWHM (km s$^{-1}$)} &
\colhead{EW (\protect\AA)} & \colhead{FWHM (km s$^{-1}$)} }
\startdata
N~\sc{iv}$\: \star$+ H~\sc{i} & 922--924;923.2 & 2.2 & 1200 & 3.1 & 1320 \\
H~\sc{i} & 926.2   & 0.18 & 340 & \nodata & \nodata             \\
H~\sc{i} & 930.7   & 0.78 & 710 & 0.44 & 240                    \\
S~\sc{vi} & 933.4 & 1.30 & 590 & 0.89 & 650                     \\
H~\sc{i} & 937.8   & 0.51 & 430 & 0.84 & 630                    \\
S~\sc{vi} & 944.5 & 1.35 & 650 & 0.81 & 820                     \\
H~\sc{i} & 949.7  & 0.88 & 630 & 1.68 & 780                     \\
N~\sc{iv}$\: \star$ & 955.3 & 0.25 & 340 & 0.21 & 320       \\
He~\sc{ii}$\: \star$ & 958.7   & 0.08 & 320 & \nodata & \nodata \\
H~\sc{i} & 972.5 & 0.82 & 530 & \nodata\tablenotemark{a} & \nodata\tablenotemark {a}     \\
C~\sc{iii} & 977.0 & 0.29 & 390 & 1.73 & 770                    \\
N~\sc{iii}$\: \star$ & 979.8 & 0.10 & 270 & 0.20 & 210      \\
N~\sc{iii} &  989.8 & 0.21 & 290 & 1.17  & 570                  \\
He~\sc{ii}$\: \star$ & 992.4   & 0.47 & 510 & 1.77  & 750   \\
Si~\sc{iii}$\: \star$ & 993.5 & 0.07 & 310 & 0.12 & 170         \\
Si~\sc{iii}$\: \star$ & 997.4 & 0.08 & 260 & 0.16 & 190         \\
N~\sc{iii}$\: \star$ & 1006.0 & 0.14 & 380 & 0.18 & 270         \\
S~\sc{iii} & 1012.5 & 0.06 & 210 & 0.09 & 320                   \\
S~\sc{iii} & 1015.5 & 0.14 & 320 & 0.21 & 530                   \\
S~\sc{iii} & 1021.1 & 0.11 & 380 & 0.27 & 430                   \\
L$\beta$  & 1025.7 & 1.27 & 440 & \nodata\tablenotemark{a} & \nodata\tablenotemark{a}   \\
O~\sc{vi} & 1031.9 & 1.46 & 850 & 1.56 & 710                    \\
O~\sc{vi} & 1037.6 & 1.71 & 840 & 1.76 & 750                    \\
S~\sc{iv} & 1062.7 & 0.37 & 410 & 0.60 & 520                    \\
Si~\sc{iv}$\: \star$ & 1066.6 & 0.24 & 540 & 0.51 & 320     \\
S~\sc{iv} & 1073.0 & 0.55 & 530 & 0.85 & 600                    \\
S~\sc{iii}$\: \star$ & 1077.2 & 0.07 & 300 & 0.05 & 290         \\
He~\sc{ii}$\: \star$ & 1084.9 & 0.41 & 490 & 3.40 & 1640    \\
S~\sc{iv}$\: \star$ & 1098.9 & 0.08 & 470 & 0.26 & 450          \\
Si~\sc{iii}$\: \star$ & 1108.4 & 0.10 & 300 & 0.83  & 620       \\
Si~\sc{iii}$\: \star$ & 1110.0 & 0.07 & 210 & 0.69 & 540    \\
Si~\sc{iii}$\: \star$ & 1113.2 & 0.20 & 340 & 0.89 & 500    \\
P~\sc{v} & 1118.0 & 0.50 & 420 & 0.19 & 350                     \\
Si~\sc{iv}$\: \star$ & 1122.5 & 0.35 & 540 & 0.61 & 380     \\
P~\sc{v}+Si~\sc{iv}$\: \star$ & 1128.0;1128.3 & 0.63 & 520 & 0.97 & 500 \\
C~\sc{iii}$\: \star$ & 1125.6 & \nodata & \nodata & 0.06 & 150  \\
Si~\sc{iii}$\: \star$ & 1144.3 & \nodata & \nodata & 0.18 & 190 \\
O~\sc{iii}$\: \star$ & 1149.6 & \nodata & \nodata & 0.10 & 160  \\
O~\sc{iii}$\: \star$ & 1150.9 & \nodata & \nodata & 0.12 & 153  \\
O~\sc{iii}$\: \star$ & 1153.8 & \nodata & \nodata & 0.10 & 140  \\
C~\sc{iv}$\: \star$ & 1169.0 & \nodata & \nodata & 0.13 & 230   \\
C~\sc{iii}$\: \star$ & 1175.3 & 0.64 & 521 & 2.20 & 650     \\
\enddata
\tablenotetext{a}{Line is blended with terrestrial airglow.}
\tablecomments{Lines marked with a $\star$ are excited state transitions.}
\end{deluxetable}

\clearpage
\begin{deluxetable}{cccc}
\tabletypesize{\scriptsize} 
\tablecaption{Prominent Interstellar
Lines in the U Gem FUV Spectrum\label{tab_ism}} \tablewidth{0pt}
\tablecolumns{4} \tablehead{ \colhead{Ion} &
\colhead{$\lambda_{lab}$ (\protect\AA)} & \colhead{$f$} & \colhead{EW
(\protect\AA)} }
\startdata
H~\sc{i} & 913.826 & 0.000170 & 0.18 \\
H~\sc{i} & 914.039 & 0.000197 & 0.18 \\
H~\sc{i} & 914.286 & 0.000229 & 0.19 \\
H~\sc{i} & 914.576 & 0.000270 & 0.17 \\
H~\sc{i} & 914.919 & 0.000321 & 0.18 \\
H~\sc{i} & 915.329 & 0.000385 & 0.17 \\
H~\sc{i} & 915.824 & 0.000468 & 0.22 \\
H~\sc{i} & 916.429 & 0.000577 & 0.18 \\
H~\sc{i} & 917.181 & 0.0007226 & 0.20 \\
H~\sc{i} & 918.129 & 0.0009213 & 0.20 \\
H~\sc{i} & 919.351 & 0.001200 & 0.20 \\
H~\sc{i} & 920.963 & 0.001605 & 0.20 \\
H~\sc{i} & 923.150 & 0.002216 & 0.21 \\
O~\sc{i} & 924.950 & 0.001540 & 0.04 \\
H~\sc{i} & 926.226 & 0.003183 & 0.22 \\
O~\sc{i} & 929.517 & 0.00229 & 0.05 \\
H~\sc{i} & 930.748 & 0.004816 & 0.22 \\
O~\sc{i} & 936.630 & 0.003650 & 0.06 \\
H~\sc{i} & 937.803 & 0.007804 & 0.23 \\
O~\sc{i} & 948.686 & 0.005420 & 0.06 \\
H~\sc{i} & 949.743 & 0.01394 & 0.24 \\
O~\sc{i} & 950.885 & 0.001570 & 0.05 \\
N~\sc{i} & 952.303 & 0.001762 & 0.02 \\
           & 952.415 & 0.001556 & 0.02 \\
N~\sc{i} & 953.415 & 0.0131  & 0.04 \\
           & 953.655 & 0.0249  & 0.05 \\
           & 953.970 & 0.0348  & 0.06 \\
           & 954.104 & 0.00675 & 0.03 \\
P~\sc{ii} & 963.801 & 1.458 & 0.03 \\
N~\sc{i} & 963.990 & 0.0148  & 0.06 \\
           & 964.626 & 0.0094  & 0.05 \\
           & 965.041 & 0.0040  & 0.03 \\
O~\sc{i} & 971.738 & 0.01367  & 0.06 \\
H~\sc{i} & 972.537 & 0.0290 & 0.27 \\
O~\sc{i} & 976.448 & 0.003310 & 0.07 \\
C~\sc{iii} & 977.020 & 0.7620   & 0.06 \\
O~\sc{i} & 988.578 & 0.0005530 & 0.18 \\
           & 988.655 & 0.008300 & \nodata\tablenotemark{a} \\
           & 988.773 & 0.04650 & \nodata\tablenotemark{a} \\
N~\sc{iii} & 989.799 & 0.1066   & 0.08 \\
Si~\sc{ii} & 989.873 & 0.1330 & \nodata\tablenotemark{a} \\
Si~\sc{ii} & 1020.699 & 0.02828 & 0.03 \\
H~\sc{i} & 1025.722 & 0.07912  & 0.44 \\
O~\sc{i} & 1025.762 & 0.01705 & \nodata\tablenotemark{a} \\
C~\sc{ii} & 1036.337 & 0.123  & 0.12 \\
O~\sc{i} & 1039.230 & 0.009200 & 0.08 \\
Ar~\sc{i} & 1048.220 & 0.263 & 0.03 \\
N~\sc{ii} & 1083.990 & 0.1031  & 0.08 \\
N~\sc{i} & 1134.165 & 0.01342 & 0.06 \\
           & 1134.415 & 0.02683 & 0.07 \\
           & 1134.980 & 0.04023 & 0.08 \\
Fe~\sc{ii} & 1144.946 & 0.15  & 0.03 \\
\enddata
\tablenotetext{a}{Feature is blended with the previous line.}
\tablecomments{Wavelengths and oscillator strengths are taken from
Morton (1991), except for \protect\ion{Fe}{2} from Howk et al.\ (2000).
The oscillator strengths of some lines have been updated by D.~C.
Morton and have been privately distributed; these were obtained
from Jenkins et al.\ (2000).}
\end{deluxetable}


\clearpage
\begin{thebibliography}{}

\bibitem[Anderson(1988)]{anderson1988} Anderson, N. 1988, ApJ, 325,
266

\bibitem[Armitage \& Livio(1998)]{armitage1998} Armitage, P.~J. \&
Livio, M. 1998, ApJ, 493, 445

\bibitem[Cheng et al.(1997)]{cheng1997}
Cheng, F.~H., Sion, E.~M., Horne, K., Hubeny, I., Huang, M. \&
Vrtilek, S.~D. 1997, AJ, 114, 1165

\bibitem[C\'{o}rdova(1995)]{cordova1995}
C\'{o}rdova, F.~A. 1995, in X-ray Binaries, eds. W.~H.~G. Lewin,
J. van Paradijs \& E.~P.~J. van den Heuvel (Cambridge:
Cambridge University Press), 331

\bibitem[C\'{o}rdova et al.(1984)]{cordova1984a}
C\'{o}rdova, F.~A., Chester, T.~J., Mason, K.~O., Kahn, S.~M. \&
Garmire, G.~P. 1984, ApJ, 278, 739

\bibitem[C\'{o}rdova \& Mason(1984)]{cordova1984b}
C\'{o}rdova, F.~A. \& Mason, K.~O. 1984, MNRAS, 206, 879


\bibitem[Cranmer \& Owocki(1996)]{cranmer1996} Cranmer, S.\ R.\ \& 
Owocki, S.\ P.\ 1996, \apj, 462, 469 

\bibitem[Drew \& Proga(2000)]{drew2000} Drew, J.\ E.\ \& Proga,
D.\ 2000, New Astronomy Review, 44, 21

\bibitem[Drew(1997)]{drew1997} Drew, J. 1997, in ASP Conf.\ Ser.\ 121:
IAU Colloq.\ 163: Accretion Phenomena and Related Outflows, 465

\bibitem[Drew(1987)]{drew1987}
Drew, J. 1987, MNRAS, 224, 595

\bibitem[Dring et al.(1997)]{dring1997} Dring, A.\ R., Linsky,
J., Murthy, J., Henry, R.\ C., Moos, W., Vidal-Madjar, A., Audouze, J., \&
Landsman, W.\ 1997, \apj, 488, 760

\bibitem[Frank, King \& Lasota(1987)]{frank1987}
Frank, J., King, A. \& Lasota, J.-P. 1987, A\&A, 178, 137


\bibitem[Harrison et al.(2000)]{harrison2000}
Harrison, T.~E., McNamara, B.~J., Szkody, P. \& Gilliland, R.~L. 2000,
AJ, 120, 2649

\bibitem[Harrison et al.(1999)]{harrison1999} Harrison, T.~E.,
McNamara, B.~J., Szkody, P., McArthur, B.~E., Benedict, G.~F.,
Klemona, A.~R. \& Gilliland, R.~L. 1999, ApJL, 515, L93 

\bibitem[Heap et al.(1978)]{heap1978}
Heap, S.~R., Boggess, A., Holm, A., Klinglesmith, D.~A., Sparks, W.,
West, D., Wu, C.~C., Boksenberg, A., Willis, A. \& Wilson, R.
1978, Nature, 275, 385

\bibitem[Hirose, Osaki, \& Mineshige (1991)]{hirose1991}
Hirose, M., Osaki, Y., \& Mineshige, S 1991, PASJ, 43, 809

\bibitem[Holm, Panek \& Schiffer(1982)]{holm1982}
Holm, A.~V., Panek, R.~J. \& Schiffer, F.~H., III 1982, ApJL, 252, L35

\bibitem[Howk et al.(2000)]{howk2000}
Howk, J.~C., Sembach, K.~R., Roth, K.~C. \& Kruk, J.~W. 2000, ApJ, 544, 867

\bibitem[Hubeny \& Lanz (1995)]{hubeny1995}
Hubeny, I., \& Lanz, T. 1995, ApJ, 439, 875

\bibitem[Hubeny, Lanz \& Jeffery(1994)]{hubeny1994}
Hubeny, I., Lanz, T. \& Jeffery, C.~S. 1994, Newsletter on
Analysis of Astronomical Spectra (St. Andrews: St. Andrews Univ.), 20, 30

\bibitem[Hubeny(1988)]{hubeny1988}
Hubeny, I. 1988, Comput. Phys. Comm., 52, 103

\bibitem[Hurwitz al.(1998)]{hurwitz1998}
Hurwitz, M., et al.\ 1998, ApJL, 500, L1

\bibitem[Hurwitz \& Bower(1996)]{hurwitz1996} Hurwitz, M. \& Bower,
S. 1996, in Astrophysics in the Extreme Ultraviolet, ed. S.~Boyer and
R.~F. Malina (Dordrecht: Kluwer Academic Publishers), 601

\bibitem[Jenkins et al.(2000)]{jenkins2000}
Jenkins, E.~B., et al.\ 2000, ApJL, 538, L81

\bibitem[Jenkins(1986)]{jenkins1986} Jenkins, E.\ B.\ 1986, \apj,
304, 739

\bibitem[Kiplinger, Sion \& Szkody(1991)]{kiplinger1991}
Kiplinger, A.\ L., Sion, E.\ M. \& Szkody, P.\ 1991, \apj, 366, 569

\bibitem[Knigge \& Drew (1997)]{knigge1997b}
Knigge, C., \& Drew, J. E. 1997, \apj, 486, 445

\bibitem[Knigge et al.(1997)]{knigge1997}
Knigge, C., Long, K.~S., Blair, W.~P. \& Wade, R.~A. 1997, \apj, 476, 291


\bibitem[Long \& Gilliland(1999)]{long1999}
Long, K.~S. \& Gilliland, R.~L. 1999, ApJ, 511, 916

\bibitem[Long \& Knigge(1998)]{long1998} Long, K.~S. \& Knigge,
C. 1998, ``Modeling the Spectral Signatures of Accretion Disk Winds in
Cataclysmic Variables'', in ``Accretion Processes in Astrophysical
Systems: Some Like It Hot'', eds. S.~S. Holt, T.~Kallman, AIP
Conference Proceedings 431, 467

\bibitem[Long et al.(1996)]{long1996}
Long, K.~S., Mauche, C.~W., Raymond, J.~C., Szkody, P. \& Mattei, J.~A.
1996, ApJ, 469, 841

\bibitem[Long et al.(1995)]{long1995} Long, K.~S., Blair, W.~P. \&
Raymond, J.~C. 1995, ApJ, 466, 964 

\bibitem[Long et al.(1994b)]{long1994b}
Long, K.~S., Wade, R.~A., Blair, W.~P., Davidsen, A.~F. \& Hubeny, I.
1994b, ApJ, 426, 704  

\bibitem[Long et al.(1994a)]{long1994a} Long, K.~S., Sion, E.~M.,
Huang, M. \& Szkody, P. 1994a, ApJL, 424, L49 

\bibitem[Long et al.(1993)]{long1993}
Long, K.~S., Blair, W.~P., Bowers, C.~W., Davidsen, A.~F., Kriss, G.~A.,
Sion, E.~M. \& Hubeny, I. 1993, ApJ, 405, 427

\bibitem[Long et al.(1991)]{long1991}
Long, K.~S., et al. 1991, ApJL, 381, L25

\bibitem[Lubow(1989)]{lubow1989}
Lubow, S.~H. 1989, ApJ, 340, 1064


\bibitem[Lynden-Bell \& Pringle(1974)]{lyndenbell1974}
Lynden-Bell, D. \& Pringle, J.~E. 1974, MNRAS, 168, 603

\bibitem[Marsh et al.(1990)]{marsh1990}
Marsh, T.~R., Horne, K., Schlegel, E.~M., Honeycutt, R.~K. \&
Kaitchuck, R.~H. 1990, ApJ, 364, 637


\bibitem[Mason et al.(1988)]{mason1988}
Mason, K.~O., C\'{o}rdova, F.~A., Watson, M.~G. \& King, A.~R. 1988,
MNRAS, 232, 779

\bibitem[Mauche(1991)]{mauche1991} Mauche, C.\ W.\ 1991, \apj, 
373, 624 

\bibitem[Mauche, Raymond \& C\'{o}rdova(1988)]{mauche1988}
Mauche, C.~W., Raymond, J.~C. \& C\'{o}rdova, F.~A. 1988, ApJ,
335, 829

\bibitem[Mauche \& Raymond(1987)]{mauche1987} Mauche, C.\ W.\ \&
Raymond, J.\ C.\ 1987, \apss, 130, 269

\bibitem[Meyer et al.(1998)]{meyer1998} 
Meyer, D.~M., Jura, M. \& Cardelli, J.~A. 1998, ApJ, 493, 222

\bibitem[Meyer et al.(1997)]{meyer1997} 
Meyer, D.~M., Cardelli, J.~A. \& Sofia, U.~J. 1997, ApJ, 490, L103

\bibitem[Meyer \& Meyer-Hofmeister(1994)]{meyer1994}
Meyer, F. \& Meyer-Hofmeister, E. 1994, A\&A, 228, 175

\bibitem[Moos et al.(2000)]{moos2000}
Moos, H.~W., et al.\ 2000, ApJL, 538, L1

\bibitem[Morton(1991)]{morton1991}
Morton, D.~C. 1991, ApJS, 77, 119

\bibitem[Morton(1978)]{morton1978}
Morton, D.~C. 1978, ApJ, 222, 863

\bibitem[Naylor \& la Dous(1997)]{naylor1997}
Naylor, T. \& la Dous, C. 1997, MNRAS, 290, 160

\bibitem[Panek \& Holm(1984)]{panek1984}
Panek, R.~J. \& Holm, A.~V. 1984, ApJ, 277, 700




\bibitem[Pringle(1988)]{pringle1988}
Pringle, J.~E. 1988, MNRAS, 230, 587

\bibitem[Prinja \& Rosen(1995)]{prinja1995} Prinja, R.\ K.\ \& 
Rosen, R.\ 1995, \mnras, 273, 461 

\bibitem[Prinja et al.(1992)]{prinja1992} Prinja, R.\ K.\ et al.\ 
1992, \apj, 390, 266 

\bibitem[Proga(1999)]{proga1999} Proga, D.\ 1999, \mnras, 304, 
938 



\bibitem[Raymond, Cox, \& Smith(1976)]{raymond1976} Raymond, J.\
C., Cox, D.\ P., \& Smith, B.\ W.\ 1976, \apj, 204, 290

\bibitem[Ritter \& Kolb(1998)]{ritter1998}
Ritter, H. \& Kolb, U. 1998, A\&AS, 129, 83

\bibitem[Sahnow et al.(2000)]{sahnow2000}
Sahnow, D.~J., et al.\ 2000, ApJL, 538, L7

\bibitem[Seaton(1979)]{seaton1979}
Seaton, M.~J. 1979, MNRAS, 187, 73P

\bibitem[Shlosman et al. (1997)]{shlosman1997}
Shlosman, I, Vitello, P., \& Mauche, C. 1997, ApJ, 471,377

\bibitem[Shlosman \& Vitello(1993)]{shlosman1993}
Shlosman, I. \& Vitello, P. 1993, ApJ, 409, 372

\bibitem[Shull et al.(2000)]{shull2000}
Shull, J.~M., et al. 2000, ApJL, 538, L73

\bibitem[Sion et al.(1998)]{sion1998}
Sion, E.~M., Cheng, F.~H., Szkody, P., Sparks, W., Gaensicke, B.,
Huang, M. \& Mattei, J. 1998, ApJ, 496, 449

\bibitem[Sion et al.(1997)]{sion1997}
Sion, E.~M., Cheng, F.~H., Szkody, P., Huang, M., Provencal, J.,
Sparks, W., Abbott, B., Hubeny, I., Mattei, J. \& Shipman, H.
1997, ApJ, 483, 907


\bibitem[Sion(1995)]{sion1995}
Sion, E.~M. 1995, ApJ, 438, 876

\bibitem[Sion et al.(1994)]{sion1994}
Sion, E.~M., Long, K.~S., Szkody, P. \& Huang, M. 1994, ApJL, 430, L53

\bibitem[Smak(1976)]{smak1976}
Smak, J. 1976, AcA, 26, 277

\bibitem[Spitzer(1978)]{spitzer1978}
Spitzer, L. 1978, Physical Processes in the Interstellar Medium (New York:
Wiley)

\bibitem[Szkody et al.(1996)]{szkody1996}
Szkody, P., Long, K.~S., Sion, E.~M. \& Raymond, J.~C. 1996, ApJ, 469, 834

\bibitem[Szkody \& Mattei(1984)]{szkody1984}
Szkody, P. \& Mattei, J. 1984, PASP, 96, 988

\bibitem[Vitello \& Shlosman(1993)]{vitello1993}
Vitello, P. \& Shlosman, I. 1993, ApJ, 410, 815

\bibitem[Wade(1984)]{wade1984}
Wade, R. A. 1984, ApJ, 335, 394

\bibitem[Wade(1981)]{wade1981}
Wade, R. A. 1981, ApJ, 246, 215

\bibitem[Warner(1995)]{warner1995}
Warner, B. 1995, Cataclysmic Variable Stars (Cambridge:Cambridge University
Press)

\bibitem[White, Nagase \& Parmar(1995)]{white1995} White, N.~E.,
Nagase, F. \& Parmar, A.~N. 1995, in X-Ray Binaries,
eds. W.~H.~G. Lewin, J. van Paradijs \& E.~P.~J. van den Heuvel,
(Cambridge: Cambridge University Press), 1

\bibitem[White \& Swank(1982)]{white1982}
White, N.~E. \& Swank, J.~H. 1982, ApJL, 253, L61

\end{thebibliography}
\end{document}